\documentclass[acmlarge, nonacm]{acmart}
\acmJournal{TOSEM}

\usepackage{graphicx}
\usepackage{tabularx} 
\usepackage{booktabs}
\usepackage{url}
\usepackage[table]{xcolor}
\usepackage{multirow}
\usepackage{makecell}
\usepackage{amsmath}
\usepackage[export]{adjustbox}  
\usepackage{tikz}
\usetikzlibrary{tikzmark}
\usetikzlibrary{arrows.meta, positioning}
\usetikzlibrary{shapes.geometric}
\usetikzlibrary{calc}
\usepackage[tikz]{mdframed}
\usepackage{enumitem}
\usepackage{listings}

\definecolor{MatplotBlue}{HTML}{0000FF}
\definecolor{MatplotRed}{HTML}{FF0000}
\definecolor{MatplotGreen}{HTML}{008000}
\definecolor{codegreen}{rgb}{0,0.6,0}
\definecolor{codegray}{rgb}{0.5,0.5,0.5}
\lstdefinestyle{mystyle}{
    commentstyle=\color{codegreen},
    keywordstyle=\color{black},
    numberstyle=\tiny\color{codegray},
    basicstyle=\ttfamily\footnotesize,
    breakatwhitespace=false,         
    breaklines=true,                 
    captionpos=b,                    
    keepspaces=true,                 
    numbers=none,                    
    numbersep=5pt,                  
    showspaces=false,                
    showstringspaces=false,
    showtabs=false,                  
    tabsize=2
}
\usepackage[textsize=tiny]{todonotes}

\newcommand{\symbolsize}{1ex}
\DeclareRobustCommand{\VarOriginal}{%
  \mbox{%
    \tikz[baseline=-0.6ex]\fill[MatplotBlue] (0,0) circle[radius=0.55\symbolsize];
    ~\textit{Original}%
  }%
}
\DeclareRobustCommand{\VarMeaningless}{%
  \mbox{%
  \tikz[baseline=-0.6ex]
    \node[regular polygon, regular polygon sides=3, draw=none, fill=MatplotGreen, minimum size=1.5\symbolsize, inner sep=0, rotate=0] {};
  ~\textit{Meaningless}%
  }%
}
\DeclareRobustCommand{\VarNoComment}{%
  \mbox{%
  \tikz[baseline=-0.6ex]
    \node[regular polygon, regular polygon sides=4, draw=none, fill=MatplotRed, minimum size=1.3\symbolsize, inner sep=0] {};
  ~\textit{NoComment}%
  }%
}

\newcommand\researchquestion[2]{
    \vspace{0.6em}
    \begin{mdframed}[
    	linewidth=1pt,
    	leftmargin=0pt,
    	topline=false,
    	rightline=false,
    	bottomline=false,
    	linecolor=gray!50]
    \begin{tabular}{@{}l m{0.89\textwidth}}
    	\textbf{{#1}} & {#2} \\
    \end{tabular}
    \end{mdframed}
    \vspace{0.3em}
}

\newcommand\RQAnswer[2]{
	\noindent 
	\begin{mdframed}[
		linewidth=1pt,
		leftmargin=0pt,
		backgroundcolor=gray!10,
		linecolor=gray!10]
		\begin{tabular}{@{}l m{0.94\textwidth}}
			\textbf{{#1}} & {#2} \\
		\end{tabular}
	\end{mdframed}
}

\newcommand\RQAnswerW[3]{
	\noindent 
	\begin{mdframed}[
		linewidth=1pt,
		leftmargin=0pt,
		backgroundcolor=gray!10,
		linecolor=gray!10]
		\begin{tabular}{@{}l m{{#1}\textwidth}}
			\textbf{{#2}} & {#3} \\
		\end{tabular}
	\end{mdframed}
}

\definecolor{pastelgreen}{HTML}{D3E4F0}
\definecolor{pastelpurple}{HTML}{D3E4F0}
\definecolor{PositiveColor}{HTML}{77DD77}
\definecolor{NegativeColor}{HTML}{FFB347}
\definecolor{LikertPurpleLight}{HTML}{e8e6f1}
\definecolor{LikertPurpleDark}{HTML}{5a3495}

\setcopyright{none}

\begin{document}

\title[A Large-Scale Experiment on Iterative Code Readability Refactoring with Large Language Models]{From Restructuring to Stabilization: A Large-Scale Experiment on Iterative Code Readability Refactoring with Large Language Models}


\author{Norman Peitek}
\affiliation{%
  \institution{Saarland University, Saarland Informatics Campus}
  \city{Saarbr{\"u}cken}
  \country{Germany} 
}

\author{Julia Hess}
\affiliation{%
  \institution{Saarland University, Saarland Informatics Campus}
  \city{Saarbr{\"u}cken}
  \country{Germany} 
}

\author{Sven Apel}
\affiliation{%
  \institution{Saarland University, Saarland Informatics Campus}
  \city{Saarbr{\"u}cken}
  \country{Germany}
}

\begin{abstract}
Large language models (LLMs) are increasingly used for automated code refactoring tasks. Although these models can quickly refactor code, the quality may exhibit inconsistencies and unpredictable behavior. In this article, we systematically study the capabilities of LLMs for code refactoring with a specific focus on improving code readability. 

We conducted a large-scale experiment using GPT5.1 with 230 Java snippets, each systematically varied and refactored regarding code readability across five iterations under three different prompting strategies. We categorized fine-grained code changes during the refactoring into implementation, syntactic, and comment-level transformations. Subsequently, we investigated the functional correctness and tested the robustness of the results with novel snippets.

Our results reveal three main insights: First, iterative code refactoring exhibits an initial phase of restructuring followed by stabilization. This convergence tendency suggests that LLMs possess an internalized understanding of an ``optimally readable'' version of code. Second, convergence patterns are fairly robust across different code variants. Third, explicit prompting toward specific readability factors slightly influences the refactoring dynamics.

These insights provide an empirical foundation for assessing the reliability of LLM-assisted code refactoring, which opens pathways for future research, including comparative analyses across models and a systematic evaluation of additional software quality dimensions in LLM-refactored code.
\end{abstract}




\maketitle

\section{Introduction}
\label{ch:introduction}

The advent of large language models (LLMs), such as GPT, Qwen, and Llama, has revolutionized the field of artificial intelligence and natural language processing. These models, powered by extensive training on diverse datasets, have demonstrated remarkable proficiency in understanding and generating human language~\cite{Myers2024}. Among the myriad applications of LLMs, their potential to enhance software development practices stands out as particularly impactful~\cite{Hou2024}.

Code quality and readability are fundamental aspects of software development~\cite{Green2011, martin_clean_2008}. Readable code not only facilitates easier maintenance and debugging but also enhances collaboration among programmers~\cite{buse_metric_2008, Oliveira2020, Borstler2023}. Code refactoring, the process of restructuring existing code to improve readability and maintainability without altering its functionality~\cite{Fowler2018}, is a critical and challenging aspect of software engineering~\cite{Almogahed2022, Peruma2022}. Poorly written code can lead to increased maintenance costs, reduced programmer productivity, and a higher likelihood of bugs and errors~\cite{Tornhill2022}. Thus, ensuring that code meets acceptable standards from its initial generation is crucial.

LLMs offer a promising avenue for both generating~\cite{Jiang2024} and refactoring code~\cite{Shirafuji2023, Cordeiro2024} with their advanced language understanding capabilities. In code generation, LLMs are able to produce initial versions of code that are clear and maintainable at first sight~\cite{Santa2025, Bistarelli2025}, hopefully reducing the need for extensive refactoring. This means that when a programmer integrates generated code into a project, it should already adhere to acceptable readability and quality standards. By leveraging its ability to comprehend and generate natural language, LLMs could assist programmers in generating or transforming complex, convoluted code written by humans or AI into more readable and maintainable versions. This would not only enhance code quality but also contribute to more efficient and effective software development processes. Given the rapid growth and adoption of LLMs, exploring their ability to refactor code is both timely and relevant. It aligns with the ongoing efforts to integrate AI into software engineering, ultimately aiming to improve productivity and code quality in the software industry~\cite{Alenezi2025, Chatterjee2024}. However, a critical question remains: Do LLMs truly have the capability to refactor code meaningfully and in a consistent way? While LLMs always generate some responses, the practicality and sensibility of these responses for code refactoring are essential questions for both research and practice. 

We conducted a large-scale experiment using OpenAI's GPT5.1 with 230 Java snippets, each systematically varied and refactored with respect to readability across five iterations under three different prompting strategies. We implemented a tool that integrates sequence-based, token-based, and AST-based similarity measures to capture fine-grained code changes, enabling the categorization of code changes into implementation, syntax, and comments. In two follow-up experiments, we investigated the functional correctness and tested the robustness of the results with snippets outside of the LLM training data.

Across conditions, our results show a consistent dynamics of ``restructuring, then stabilizing'': Early iterations make substantial changes (e.g., identifier renaming, decomposition into more methods, comment pruning), followed by iterations that converge toward a stable representation that appears to reflect an internalized notion of readable code. This convergence holds even when inputs are purposefully degraded (meaningless identifiers or removed comments), though the changes differ by starting point. Targeted prompts steer the types of changes without altering the overall convergence trend, and naming-focused prompts may induce oscillatory renaming behavior. Functionality breaks due to the refactoring are rare but non-zero with each iteration. Our follow-up experiments further support the robustness of our main findings.

These results suggest LLMs can reliably normalize diverse code snippets toward consistent coding style, but they also highlight the need for practical guardrails. Usage of automated refactoring should use explicit stopping criteria to curb over-refactoring, and prompt phrasing must be considered carefully to avoid rename oscillations and mechanisms to preserve (for humans) valuable explanatory comments. In addition to our empirical findings, our LLM-agnostic framework enables exact and non-exact replications and comparative studies across models. To this end, we support further systematic assessment of additional quality attributes (e.g., code smells or complexity metrics) and pave the way for exploring long-term evolutionary dynamics in AI-refactored code.

In summary, we make the following contributions:
\begin{itemize}
    \item Empirical evidence that LLM-based readability refactoring tend toward convergence, while on occasion remaining vulnerable to back-and-forth changes, highlighting both their strengths and limitations.
    \item Insights into how explicit prompt strategies influence refactoring trajectories, emphasizing the need for careful prompt engineering.
    \item A reusable, LLM-agnostic framework for examining code evolution under iterative refactoring, which can apply similar research goals to various LLMs beyond GPT5.1.
    \item An online replication package\footnote{\url{https://github.com/brains-on-code/IterativeRefactoringLLM}}, including all snippets, data, additional figures, and scripts for the full framework.
\end{itemize}

The remainder of this paper is structured as follows: Section \ref{ch:background} provides the necessary background, including fundamental concepts relevant to this research, and discusses related work, summarizing existing approaches and highlighting the gap this experiment aims to address. Section \ref{ch:methodology} presents the methodology of our main experiment. Section \ref{ch:results} reports the findings of our experiment, while Section \ref{ch:discussion} provides a critical discussion. Section \ref{ch:threats} presents two small-scale robustness analyses for some key design decisions, outlines open issues, and elaborates on further threats to validity. Finally, Section \ref{ch:conclusion} concludes the paper.

\section{Background and Related Work}
\label{ch:background}

The rapid advancement of artificial intelligence (AI) and large language models (LLMs) has had a profound impact on software engineering, particularly in the areas of program comprehension, readability assessment, and automated code generation. In this section, we provide the necessary theoretical and technical foundations for our study discussing the relevant key concepts.

\subsection{Program Comprehension}

An understanding of source code is crucial in software engineering, as it directly impacts tasks such as debugging, code review, and refactoring. The term \emph{program comprehension} refers to the cognitive processes involved in understanding code, a fundamental task in software engineering~\cite{Siegmund2016}. Program comprehension involves cognitive processes, such as pattern recognition, memory retrieval, and abstraction. Research in this area seeks to determine what factors influence comprehension and how programmers interact with code on a cognitive level~\cite{wyrich_40_2024}.

Early studies primarily relied on self-reports and observational methods to assess comprehension difficulty~\cite{tiarks_what_nodate}. More recent approaches integrate eye-tracking and neuroimaging techniques (e.g., fMRI, EEG) to gain insights into how programmers process code at a neuro-cognitive level~\cite{siegmund_understanding_2014, peitek_simultaneous_2018, Gonccales:2021, fakhoury_measuring_2020}. These studies suggest that expert programmers rely more on top-down comprehension, forming hypotheses about a code snippet before verifying details, while novices adopt a bottom-up approach, building an understanding incrementally \cite{siegmund_measuring_2017}.

\subsection{Code Readability Models}

A crucial aspect of program comprehension is readability, referring to the ease with which code can be understood. Readability is influenced by aspects, such as variable naming, indentation, comments, and syntactic simplicity \cite{martin_clean_2008, buse_metric_2008}. The development of readability models aims to quantify and predict how readable a given piece of code~is~\cite{buse_learning_2010, scalabrino_comprehensive_2018}.

Traditional code readability models are based on heuristic metrics, such as Halstead complexity measures and McCabe’s Cyclomatic Complexity~\cite{halstead_elements_1977}. While these models provide quantifiable measures of code complexity, they often fail to capture the subjective and context-dependent nature of readability~\cite{Peitek2021}. Readability is not solely determined by syntactic complexity but also influenced by individual cognitive factors, programming experience, and familiarity with coding conventions~\cite{wagner_code_2021}. As a result, purely heuristic approaches may misrepresent how programmers actually perceive and evaluate code readability. To address these limitations, machine learning (ML) approaches have been introduced~\cite{scalabrino_automatically_2021, vitale_using_2023}. 

To predict readability scores, ML-based readability models not only leverage hand-crafted features, such as line length, identifier complexity, and indentation consistency, but also leverage large datasets and empirical readability assessments to provide more accurate and context-aware predictions. More recently, deep learning methods, particularly transformer-based models such as \texttt{CodeBERT} and \texttt{GraphCodeBERT}, have outperformed previous approaches by learning semantic representations from large-scale code corpora~\cite{feng_codebert_2020}. These models integrate syntactic structure with semantic context, improving their ability to assess code readability.

Beyond static analysis, readability models are now being adapted into integrated development environments, offering real-time feedback to programmers. Additionally, integrating LLMs for readability assessment is an emerging trend, as these models can refactor code suggestions dynamically based on human feedback~\cite{guo_exploring_2024}.

\subsection{Large Language Models in Software Engineering}

The field of natural language processing has undergone a rapid transformation with the advent of LLMs. Early language models relied on n-gram and statistical methods, gradually evolving into recurrent neural networks and long short-term memory networks~\cite{bengio_neural_nodate}. The paradigm shift occurred with the introduction of the transformer architecture, which laid the foundation for models such as \texttt{GPT} and \texttt{BERT}~\cite{vaswani_attention_nodate}.

Zheng et al. provide an extensive investigation into the integration of LLMs in software engineering, categorizing and analyzing 123 relevant research papers~\cite{zheng_towards_2023}. They emphasize the growing importance of LLMs in various software engineering tasks, including code generation, vulnerability detection, and program repair, highlighting both the potential and current limitations of these models. They note that LLMs excel in tasks requiring syntactical understanding, such as code summarization, but LLMs struggle with tasks requiring deep semantic comprehension, such as complex code generation and vulnerability detection~\cite{zheng_towards_2023}. In the same vein, Hou et al. conducted a systematic review, focusing on the broader application of LLMs in software engineering~\cite{Hou2024}. They also find an increasing integration of LLMs in software engineering, particularly noting the dominance of decoder-only models in tasks requiring code generation and completion. Furthermore, they identified key challenges, such as the need for better dataset curation and the development of more sophisticated evaluation metrics, to enhance the effectiveness of LLMs in software engineering~\cite{Hou2024}.

Clearly, there is the need for further refinement in LLMs, particularly in improving their understanding of code semantics and reliability in more complex tasks. As LLMs continue to improve, their potential to support program comprehension extends beyond textual explanations~\cite{zheng_towards_2023, Hou2024}. This growing role of LLMs in program comprehension underscores the need for a deeper investigation into their effectiveness and limitations, particularly in the context of automated refactoring. In the context of our experiment, it is important to situate this concept within the broader framework program comprehension, as it represents a fundamental skill for interpreting and working with code, which is directly relevant to refactoring, and therefore forms the foundation for the refactoring tasks performed by LLMs.

\subsubsection{Code Generation with Large Language Models}

Tian et al. conducted an empirical study to assess ChatGPT's potential as a fully automated programming assistant, focusing on three core tasks: code generation, program repair, and code summarization~\cite{tian_is_2023}. The study revealed that, while ChatGPT excels in generating correct code for common programming problems, it struggles with generalizing to new, unseen challenges. Additionally, the study highlighted that ChatGPT's performance in program repair is competitive but limited by its attention span and the effectiveness of provided problem descriptions. The model demonstrated an unexpected capability in summarizing code, but its current limitations in handling novel problems and providing consistent repair solutions indicate that further refinement is necessary. Similarly, Jin et al.~\cite{jin_can_2024} conducted an empirical evaluation of ChatGPT's effectiveness in supporting programmers, focusing on code generation. Their study analyzed interactions from the DevGPT dataset, comprising real-world programmer conversations involving ChatGPT. The findings reveal that, while ChatGPT is frequently used for generating code, its output is typically more suited for demonstrating concepts or providing documentation examples, rather than being production-ready. The study highlights that generated code often requires significant modifications before integration into production, and the tool is most effective in contexts where programmers request improvements or additional context rather than entirely new code generation. This research underscores the current limitations of LLMs in practical software development and emphasizes the need for further refinement before LLMs can be fully integrated into modern development workflows. 

With a more technical perspective, Liu et al. empirically assessed ChatGPT's code generation capabilities~\cite{liu_no_2024}. Their findings reveal that ChatGPT is effective at generating functionally correct code for problems dated before 2021. However, its performance significantly drops for more recent problems. Notably, Liu et al. highlight the limitations of ChatGPT's multi-iteration fixing process, where attempts to correct erroneous code often result in increased code complexity without fully resolving functional issues. Similarly, Liu et al. investigated the quality and reliability of code generated by ChatGPT~\cite{liu_refining_2023} and highlight substantial challenges in the correctness and maintainability of ChatGPT-generated code. Nearly half of the generated code suffers from maintainability issues, such as poor code style and unnecessary complexity. Moreover, while ChatGPT can address some of these issues when provided with specific feedback, the model frequently introduces new problems during the refactoring.

\subsubsection{Code Refactoring with Large Language Models}

Martinez et al. reviewed 50 relevant research papers on code refactoring with LLMs~\cite{Martinez2025}. They found a large variety of approaches, programming languages, prompting techniques, and ways to measure different refactoring outcomes. Specifically, it appears difficult for researchers to objectively assess whether refactored code is actually preferable. Thus, researchers have relied on indirect assessments, such as whether the refactored version has fewer code smells~\cite{Cordeiro2024} or lower code complexity metrics, such as cyclomatic complexity~\cite{Choi2024}, in relying on human judgment~\cite{Chand2025}.

One approach to address code quality issues of LLMs is using their self-verification capabilities (e.g., self-reflection with critique-and-repair) . Yu et al. explored this aspect with ChatGPT, but found limitations in its self-verification process~\cite{yu_fight_2024}. This includes a high rate of erroneous self-assessments, where the model often incorrectly predicts the correctness, security, and success of its code outputs. This is particularly relevant to refactoring, where programmers may be able to increase the quality of the responses with targeted prompts. AlOmar et al. conducted an exploratory study examining programmer-LLM interactions during code refactoring tasks~\cite{alomar_how_2024}. Using the DevGPT dataset, they found that programmers frequently use generic terms when requesting refactoring, while LLMs often explicitly state the refactoring intention, focusing on improving quality attributes (e.g., maintainability, readability). Similarly, DePalma et al. conducted an empirical study to assess LLMs capabilities in performing code refactoring~\cite{depalma_exploring_2024}. The study involved refactoring 40 Java code segments across eight quality attributes, including performance, complexity, and readability. The results indicate that LLMs were successful in refactoring code in 319 out of 320 trials, offering both minor and significant improvements. However, LLMs struggled with complex refactoring tasks, often applying superficial changes, though, that did not address deeper issues. A recent benchmark evaluating LLMs on repository-level refactorings showed a low success rate~\cite{Xu2026}. This provides further evidence that LLMs can be a valuable tool for refactoring tasks, but we must carefully investigate their contextual strength and weaknesses.

Guo et al. present an empirical study examining ChatGPT’s potential for automated code refactoring in the context of code review~\cite{guo_exploring_2024}. Using the CodeReview benchmark and a newly constructed high-quality dataset, they compare GPT-3.5 and GPT4 against CodeReviewer, a state-of-the-art tool based on CodeT5. Their findings show that GPT achieves superior generalization and higher EM and BLEU scores than CodeReviewer on the new dataset, although performance remains limited overall. Notably, they highlight that GPT performs best with low temperature\footnote{Temperature is a parameter during the text generation process that controls the randomness of the LLM output. A lower value moves the LLM to a more conservative and deterministic output, while a high value makes it more unexpected or creative.} settings and concise, scenario-based prompts.

Hu et al. investigate the robustness of LLMs when confronted with poorly readable code, a dimension largely neglected in existing evaluation benchmarks~\cite{hu_how_2024}. While prior research has primarily tested LLMs on well-structured, high-readability code, this study systematically degrades readability through obfuscation techniques that perturb both semantic (e.g., identifiers, function names) and syntactic (e.g., operators, branches) features. Their empirical results demonstrate that current models exhibit a strong dependency on semantic cues and perform poorly when these cues are obscured, revealing limited robustness to syntactic variations. This highlights a critical gap of LLMs in real-world scenarios, where code readability often varies significantly.

To address these gaps, Liu et al. propose CodeQUEST, a framework that leverages GPT-4o for iterative code quality evaluation and enhancement across multiple dimensions (including readability, maintainability, testability, efficiency, and security)~ \cite{liu_iterative_2025}. The framework integrates an evaluator, which provides structured quantitative and qualitative assessments, with an optimizer that applies the feedback to refactor code in successive cycles. Evaluated on 42 Python and JavaScript examples, CodeQUEST achieved improvements in 41 cases, with the majority of gains occurring in early iterations, and demonstrated stronger alignment with established metrics such as Pylint, Radon, and Bandit compared to a baseline. Notably, the framework was able to identify issues overlooked by traditional tools, particularly in security and scalability, highlighting the potential of LLMs for systematic and multi-faceted code refactoring while acknowledging limitations regarding subjectivity, stochasticity, and language coverage.

Overall, these findings underscore (1) the need for human oversight and the development of more sophisticated verification mechanisms before LLMs can be trusted fully; (2) programmers need to craft more specific prompts to maximize the effectiveness of LLMs in refactoring tasks; and (3) we must investigate the role of LLMs in improving code quality and readability, particularly in the context of automated refactoring.

\subsection{Research Gap Analysis}
\label{sec:gap-analysis}

LLMs have shown potential in automating code refactoring and improvement tasks, yet fundamental questions about their behavior under iterative refactoring, especially with regards to readability, remain unresolved. While prior work has raised concerns about ``reality distortion'' and contradictory improvements in AI-generated content~\cite{shumailov_ai_2024}, little is known about what actually happens when LLMs are repeatedly prompted to refactor their own code outputs. Specifically, it is unclear whether these models eventually converge toward stable, higher-quality solutions (potentially guided by implicit notions of best practices) or whether they continue to introduce superficial, oscillating, or even regressive changes across iterations. In the same vein, the degree to which such iterative refactoring differ under unguided versus targeted prompts, or when applied to code of varying initial quality, has not been systematically investigated. Existing studies acknowledge the ability of LLMs to refactor code~\cite{depalma_exploring_2024}, but they stop short of analyzing how the nature of changes evolve over several refactoring iterations. 

To address this gap, we developed a framework for systematically analyzing code changes produced by LLMs across multiple iterations. Using this framework, we conducted a large-scale study that applies multiple refactoring iterations to an existing set of code snippet. The resulting data are analyzed with respect to how changes emerge at different refactoring stages, and then progressively decomposes these changes by type (i.e., insertions, deletions, and changes of code elements). This layered analysis enables a deeper understanding of the patterns underlying iterative refactoring, providing insights into whether and how LLMs move toward convergence, and under which prompting or code-quality conditions such processes are more or less effective.
\section{Methodology}
\label{ch:methodology}

Our main experiment aims to analyze how LLMs change source code across multiple iterations with the goal of improving code readability, focusing on measurable patterns of change. To this end, we conducted a systematic, large-scale analysis using OpenAI's GPT5.1 with a structured dataset and quantitative metrics. Our main experiment addresses three research questions, which we introduce next.

\subsection{Research Questions and Operationalization}
\label{subsec:research-questions}

We aim to answer three research questions. Each of them addresses a different dimension of the role of LLMs in iterative code refactoring and is motivated by gaps in prior research and the objectives of this experiment. For RQ\textsubscript{1} and RQ\textsubscript{2}, our aims are content-driven and operationalized using the metrics described in Section~\ref{subsec:metrics}. For RQ\textsubscript{3}, we pose a hypothesis, which is directly testable with statistical methods.

\subsubsection{Evolution of Iterative Refactoring (RQ\textsubscript{1})} 

\mbox{} 
\vspace{0.5em}

\researchquestion{RQ\textsubscript{1}}{How do iterative refactoring by LLMs evolve when provided with a code snippet that already adheres to best practices for code readability?}

This question is concerned with whether LLMs are able to preserve high-quality code without introducing unnecessary or contradictory changes. It is crucial because a refactoring assistant should ideally recognize when no further improvement is required and avoid degrading code quality.

\paragraph{Operationalization} 

We measure this by tracking the number and types of changes (code implementation changes including access, call, control, literal, operator, and other structural changes; syntax-only changes; renames; comment changes; mixed changes), and structural stability (total lines, code lines, comment lines, inline comments, empty lines, number of methods).

\subsubsection{Convergence Across Code Variants (RQ\textsubscript{2})} 

\mbox{} 
\vspace{0.5em}

\researchquestion{RQ\textsubscript{2}}{When multiple variations of the same code snippet---each changed with respect to a code readability aspect---are independently of each other iteratively refactored, do they converge after a certain number of iterations?}

This question investigates whether LLMs normalize different starting points into similar final snippets, which would indicate a consistent understanding of coding conventions. Convergence is of particular interest, as it reflects the stability and reliability of the refactoring across diverse inputs.

\paragraph{Operationalization} 

We assess convergence using the proportion of unchanged code lines, average similarity scores of changed lines, and structural correspondence from the absolute values metrics. We track the total insertions and deletions per iteration to discover stabilization trends, and cross-variant comparisons reveal whether final code reflects convergence or retains traces of the original variant.

\subsubsection{Impact of Explicitly Emphasizing Readability Aspects (RQ\textsubscript{3})} 

\mbox{} 
\vspace{0.5em}

\researchquestion{RQ\textsubscript{3}}{Do targeted refactoring become more effective when the prompt explicitly emphasizes a readability aspect?}

This question addresses the influence of prompting on refactoring quality. Since prior research has shown that LLMs are highly sensitive to input phrasing~\cite{Qian2024, Ismithdeen2025}, we assess whether explicitly guiding the LLM toward a specific readability aspect (e.g., identifier naming, comments) leads to more consistent and meaningful refactoring compared to unguided prompts.

\paragraph{Operationalization} 

We measure the effectiveness of different prompts with the same metrics as described above, with the additional hypothesis that it increases the alignment with the emphasized aspect, and a faster stabilization of the refactored code across iterations.

\subsection{Snippet Sampling and Data Preprocessing}
\label{subsec:snippet-sampling}

To ensure the suitability and comparability of code snippets, we decided on four criteria:
 
 \begin{enumerate}
 	\item \textbf{LOC Range:} The file has between 50--200 lines (i.e. overall size of the file).
 	\item \textbf{Methods Ratio} The chosen files must have a comparable structural complexity. This is measured by number of methods given the file size (i.e. 1--3 methods per 50 lines).
 	\item \textbf{Code/Comments Ratio:} A file must contain a significant amount of code as compared to comments ($\geq$ 50\% code lines).
 	\item \textbf{Best Practices:} All snippets should follow a comparable standard of code quality.
 \end{enumerate}

\begin{table}[ht]
	\centering
    \caption{Distribution of \texttt{.java} files across LOC intervals. 
	\#~Files lists all files per interval, \#~Method Ratio and \#~Code/Comments Ratio indicate those meeting the respective conditions, and \#~Total Valid Files shows files satisfying both.}
	\label{tab:snippet-sampling-criteria}
    \begin{tabular}{rrrrr}
        \toprule
        LOC Interval & \# Files & \# Methods Ratio & \# Code/Comments Ratio & \# Valid Files \\ 
        \midrule
        0--49 & 209 & 202 & 150 & 143 \\
        50--99 & 287 & 227 & 187 & 148 \\
        100--149 & 75 & 56 & 60 & 45 \\
        150--199 & 45 & 32 & 38 & 28 \\
        200--249 & 18 & 9 & 17 & 9 \\
        $>$ 250 & 24 & 13 & 4 & 3 \\
        \bottomrule
    \end{tabular}
\end{table}

For our experiment, we used the Java code files from the GitHub repository \texttt{``The Algorithms – Java''}~\footnote{\url{https://github.com/TheAlgorithms/Java/}} for three reasons: (1) This repository provides a structured and diverse collection of Java implementations of various algorithms, covering a wide range of problem domains such as searching, sorting, mathematics, and data structures; (2) it is designed for educational and comparative purposes, ensuring that the code follows consistent formatting and best practices, which facilitates systematic analysis; (3) as an open-source repository under MIT license with active contributions, it ensures accessibility and reproducibility, both of which are essential for a rigorous research experiment.

We summarized the counts of files in the repository that fulfill our conditions (excluding test files) in Table~\ref{tab:snippet-sampling-criteria}. There are $230$ of originally $658$ files that fulfill all of our criteria. Next, for each of these $230$ files, we created three variants:
\begin{itemize}
	\item \textbf{\VarOriginal~variant}: The unchanged original implementation.
	\item \textbf{\VarMeaningless~variant}: Variable, method, and class names as well as all content in JavaDocs and comments are intentionally made meaningless, while preserving structure.
	\item \textbf{\VarNoComment~variant}: All JavaDocs as well as block and inline comments are completely removed.
\end{itemize}

\begin{table}[ht]
	\centering
    \caption{Wording of the three different prompts used in the experiment.}
	\label{tab:prompts}
	\begin{tabular}{llp{10cm}}
		\toprule
		Prompt ID & Prompt Type & Exact Prompt Wording \\ 
        \midrule
		  Prompt\textsubscript{General} & Unguided & \textit{``Refactor this code for improved readability.''} \\ 
        Prompt\textsubscript{Meaning} & Targeted & \textit{``Refactor this code for improved readability, especially with respect to identifier naming.''} \\ 
        Prompt\textsubscript{Comments} & Targeted & \textit{``Refactor this code for improved readability, especially with respect to comments.''} \\
        \bottomrule
	\end{tabular}
\end{table}

To answer RQ\textsubscript{1} and RQ\textsubscript{2}, we used the same Prompt\textsubscript{General} for iterative refactoring (i.e. \textit{version creation}), while we used two adapted prompts (Prompt\textsubscript{Meaning}, Prompt\textsubscript{Comments}) to answer RQ\textsubscript{3} (effect of explicit prompting). Table~\ref{tab:prompts} provides an overview of the used prompt strategies.

\begin{figure}[ht!]
	\centering
	\includegraphics[width=\textwidth, trim=0cm 1.3cm 0.5cm 0cm, clip]{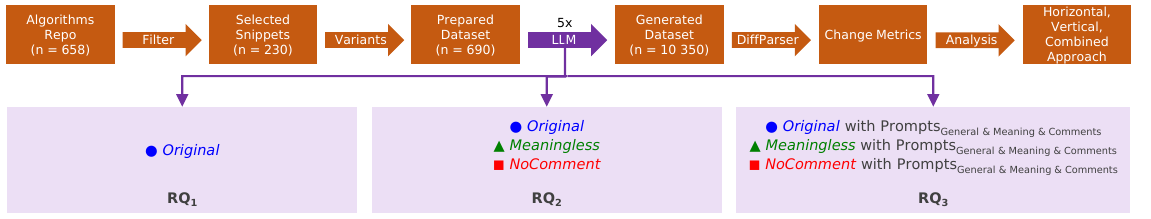}
	\caption{Overview of the experiment flow from source data to data analysis.}
    \Description{TODO}
	\label{fig:experiment-overview}
\end{figure}

To scale the experiment, we used the OpenAI API to generate outputs in an automated and controlled manner. We configured the API call with \texttt{temperature $= 0$} to minimize non-deterministic output variations, as recommended in prior research~\cite{ouyang_empirical_2025}. We used the full \texttt{gpt-5.1} model. Our implementation and interaction with the OpenAI API is stateless, which means each request is independent of previous ones to eliminate a bias from prior context. We illustrate the full experiment flow in Figure~\ref{fig:experiment-overview}.

\paragraph{Summary of Dataset Creation}

In total, we selected $230$ Java files for our main experiment. Each file was extended into two additional variants, resulting in three variants per original file. Every variant was then iteratively refactored across five iterations using three distinct prompt strategies. Thus, we generated a total of $10\,350$ snippets ($230$ snippets $\times\,\,3$ variants $\times,\,3$ prompts $\times,\,5$ iterations).

\subsubsection{DiffParser: A Tool for Tracking Code Changes}
\label{subsec:diffparser}

To conduct our experiment and evaluate the code comparisons based on the metrics (described in the subsequent Section~\ref{subsec:metrics}), we developed a set of Python scripts tailored to our experiment, including a custom \emph{DiffParser} designed to meet our specific requirements. The resulting output consists of three components:  
\begin{enumerate}
    \item a mapping of changes (removed-added line pairs), 
    \item a list of deletions (removed lines without a match), and  
    \item a list of insertions (added lines without a match).  
\end{enumerate}

Together, these outputs provide a complete classification of all line-level changes between the two code versions. The technical explanation would go beyond the scope of this paper and is available including the complete implementation in the online repository.

\subsection{Metrics}
\label{subsec:metrics}

To systematically evaluate the iterative refactoring produced by the LLM, we employ a set of metrics that capture both the nature and progression of code changes. The analysis is designed to answer the research questions by assessing different aspects of code evolution, stability, and convergence.

\subsubsection{Absolute Values}

\begin{table}[ht]
	\centering
    \caption{Absolute metric values for the exemplary snippet \texttt{MaximumSumOfNonAdjacentElements} in \VarOriginal.}
	\label{tab:anagrams-absolute-values-metrics}
	\resizebox{\textwidth}{!}{
		\begin{tabular}{llrrrrrr}
			\toprule
			\textbf{Version} & \textbf{Prompt} & \textbf{Total Lines} & \textbf{Code Lines} & \textbf{Comment Lines} & \textbf{Inline Comments} & \textbf{Empty Lines} & \textbf{Methods} \\ 
            \midrule
			0 & ---  & 95 & 41 & 35 & 3  & 19 & 2 \\
			1 & Prompt\textsubscript{General} & 82 & 42 & 27 & 0 & 13 & 2 \\ 
			1 & Prompt\textsubscript{Meaning} & 81 & 41 & 23 & 0 & 17 & 2 \\
			1 & Prompt\textsubscript{Comments} & 88 & 40 & 36 & 0 & 12 & 2 \\
			2 & Prompt\textsubscript{General} & 90 & 47 & 27 & 0  & 16 & 4 \\
			2 & Prompt\textsubscript{Meaning} & 79 & 41 & 21 & 0  & 17 & 2 \\
			2 & Prompt\textsubscript{Comments} & 94 & 40 & 42 & 8 & 12 & 2 \\
            \bottomrule
		\end{tabular}
	}
\end{table}

For each snippet instance (identified as a unique combination of snippet, variant, version, and prompt), we compute a set of absolute metrics to characterize its structural properties. These metrics include the total number of lines, lines of code, comment lines, inline comments, empty lines, and number of methods. We show a partial example for the resulting data in Table~\ref{tab:anagrams-absolute-values-metrics}.

\subsubsection{Comparison Values}  

For each comparison of snippet instances (identified as a unique combination of snippet, variant 1, variant 2, version 1, version 2, and prompt) is analyzed with respect to the following set of comparison values:

\paragraph{Unchanged Code}

The unchanged code part represents the number of lines that remain identical across two code versions. Conceptually, each version of the code can be decomposed into three disjoint categories: unchanged lines, changed lines, and either insertions (for the new version) or deletions (for the old version). By construction, both quantities are equal, yielding the unchanged code part as a consistent measure of preserved code between the two versions.

\paragraph{Changed Code}

The analyzed code changes are categorized into distinct change types to systematically capture different aspects of code evolution.

\begin{itemize}
    \item \textbf{Rename:} Changes involving consistent renaming of identifiers, such as variables, methods, or constants, without affecting program behavior.
    \item \textbf{SyntaxOnly:} Purely syntactic adjustments, including formatting changes, reordering of code segments, or other non-semantic edits that do not alter program execution.
    \item \textbf{CommentChange:} Changes restricted to comments or documentation, capturing updates in explanatory or descriptive content without altering code logic.
    \item \textbf{MixedChange:} Instances where multiple types of changes occur simultaneously within a code line, reflecting combinations of semantic, syntactic, or identifier-level changes.
    \item \textbf{CodeChange:} Aggregates subcategories of behavioral or functional changes:
    \begin{itemize}
        \item \textbf{AccessChange:} Changes to data structure or collection accesses.
        \item \textbf{CallChange:} Changes to function or method calls.
        \item \textbf{ControlChange:} Adjustments to control-flow statements such as conditionals or loops.
        \item \textbf{LiteralChange:} Changes to constant values.
        \item \textbf{OperatorChange:} Changes in operators affecting computation.
        \item \textbf{OtherStructuralChange:} Alterations in structural constructs (e.g., method signatures or data structures).
    \end{itemize}
\end{itemize}

We focus on on these types because together they isolate different aspects of code readability. Renaming, SyntaxOnly, and CommentChange capture three core non-behavioral dimensions of readability (lexical semantics of identifiers, surface formatting, and expressing intent). Furthermore, they map to our directed prompt, which allows us to measure them separately reducing potential confounds. MixedChange explicitly measures co-occurring changes within a line, which is fairly common, to avoid misclassifying composite readability actions. CodeChanges are tracked to capture functional changes that may accompany readability refactorings.

\paragraph{Average Similarity Score}

To quantify the degree of resemblance between changed lines across two code versions, we implemented an average similarity score $\overline{\operatorname{sim}}$. We defined it by first measuring the similarity sim$(l_i, l'_i)$ of each changed line $l_i$ with its corresponding line $l'_i$ in the preceding version, and then computing the arithmetic mean across all $n$ changed lines:

	\[
	\overline{\operatorname{sim}} = \frac{1}{n} \sum_{i=1}^{n} \text{sim}(l_i, l'_i).
	\]
    
This aggregated metric provides an overall measure of how similar the changed code segments remain between two snippets.

\paragraph{Total Insertions}

Total number of insertions measures the total number of lines added in each snippet, further distinguished by type: code lines, comment lines, and empty lines.

\paragraph{Total Deletions}

Total number of deletions measures the total number of lines removed in each snippet, similarly categorized into code lines, comment lines, and empty lines.
	
\paragraph{Relative Changes}

Using these metrics, we further quantify the relative changes in absolute values, such as total lines, code lines, comment lines, inline comments, empty lines, and number of methods, across snippet comparisons.

\vspace{1em}

The online repository contains one representative source-target pair per classification to illustrate the types of changes. It is important to note that comparisons are only conducted between snippets sharing the same name. Depending on the analysis, we either compare snippets of the same variant across different versions (horizontal comparison) or snippets of the same version across different variants (vertical comparison), which we will explain in more detail next.


\subsection{Comparison Strategies for Measuring Refactoring Trajectories}
\label{subsec:analysis-approaches}

\begin{figure}[ht!]
	\centering
	\includegraphics[width=0.8\textwidth]{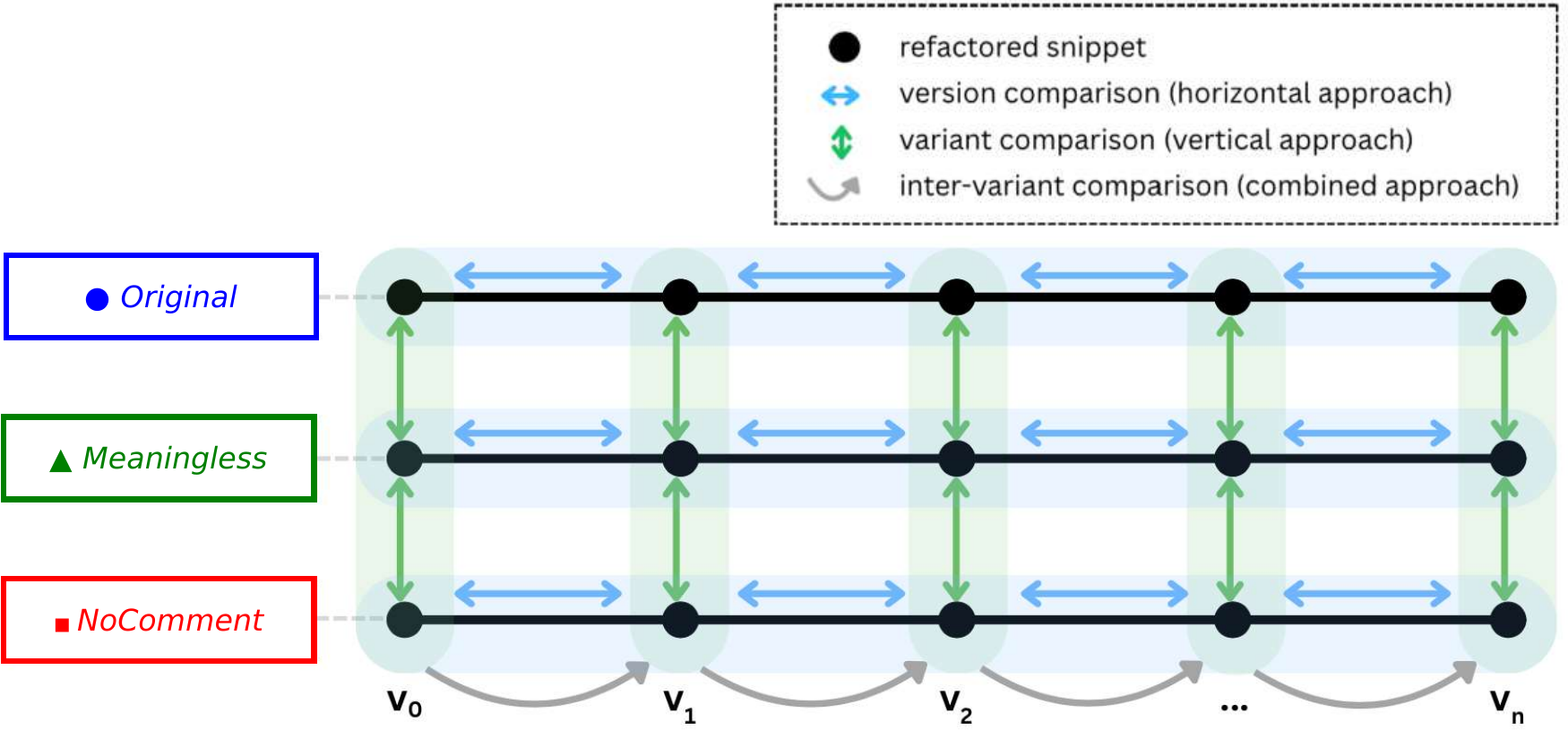}
	\caption{Overview of the three types of comparison (non-consecutive comparisons are omitted for readability)}
    \Description{TODO}
	\label{fig:horizontal-vertical-approaches}
\end{figure}

Understanding how code evolves over multiple iterations is essential for assessing the consistency and effectiveness of iterative refactoring. In the context of our experiment, we investigate whether LLM-based refactoring follows a coherent progression, whether they exhibit signs of inconsistency (e.g., back-and-forth changes), and whether different refactoring paths converge toward a common structure. To systematically analyze these aspects, we consider three complementary approaches: Horizontal approach, vertical approach, and combined approach (cf. Figure~\ref{fig:horizontal-vertical-approaches}).

\subsubsection{Horizontal Comparison} 

\begin{figure}[ht!]
	\centering
	\includegraphics[width=0.3\textwidth]{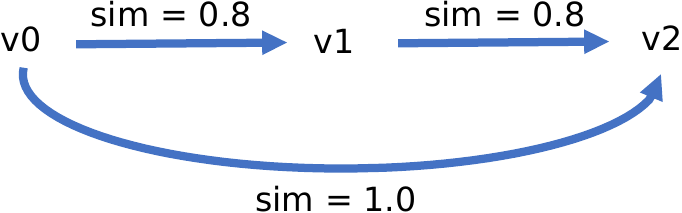}
	\caption{Schematic visualization of similarity score evolution in back-and-forth changes with v0 $=$ v2 $\neq$ v1}
    \Description{TODO}
	\label{fig:back-and-forth}
\end{figure}

We compare different versions of the same variant over time. Instead of only considering directly consecutive versions, we compare each version with \textit{all} preceding ones. This allows us to track how a particular variant evolves through iterative refactoring, highlighting the nature and consistency of changes applied by the LLM across different versions. If versions that are further apart show higher similarity than those that are closer together, this may indicate back-and-forth changes. 
	
Figure~\ref{fig:back-and-forth} illustrates the importance of comparing non-consecutive versions. Let v0, v1, and v2 denote three consecutive versions of a snippet. The LLM introduces changes from v0 to v1, which are subsequently reversed from v1 to v2. As a result, the similarity between consecutive versions is $\text{sim}(v0, v1) = \text{sim}(v1, v2) = s < 1$, while the non-consecutive versions are identical, $\text{sim}(v0, v2) = 1$. Comparing only consecutive versions would therefore fail to detect the equivalence between v0 and v2.

\subsubsection{Vertical Comparison} 

In a vertical comparison, code variants are compared within the same version number. This enables an analysis of how different refactoring for the same base code differ at a given stage, revealing potential diversity in the refactoring strategies generated by the LLM.

\subsubsection{Combined Comparison} 

A combined comparison investigates whether different code variants gradually converge toward an ``optimal variant'' as the refactoring progresses (i.e., later iterations). Analyzing both horizontal and vertical dimensions provides insights into whether the refactoring leads to a consensus on best practices or whether structural differences persist over multiple iterations.

\section{Results}
\label{ch:results}

This section presents the results of our experiment following the methodology outlined in Section~\ref{ch:methodology}. First, we report on the absolute structural metrics to provide a descriptive baseline of the code snippets across refactoring iterations. Subsequently, we analyze various aspects of code evolution, including the nature and distribution of code changes, as well as trends in structural convergence and stability. The presentation of the results are organized according to our research questions.

\subsection{Evolution of Iterative Refactoring (RQ\textsubscript{1})}
\label{sec:RQ1}

To address RQ\textsubscript{1}, we investigate how \texttt{GPT5.1} behaves when asked to improve the readability of code that requires little to no improvement. This setting allows us to examine whether LLMs merely preserve well-structured input, introduce unnecessary code changes, or converges toward a stable representation over multiple refactoring iterations. To this end, we focus on the original code snippets (), which based on the educational goals of the dataset aim to follow established best practices in identifier naming, commenting, and formatting. Each snippet is iteratively refactored across five iterations using Prompt\textsubscript{General} (\textit{``Refactor this code for improved readability.''}). Analyzing these five iterations enables us to study the refactoring dynamics of LLMs when no substantial restructuring is required and to determine whether stability or persistent micro-changes dominate over time.

\subsubsection{Absolute Code Metrics Across Iterations (\VarOriginal{})}

\begin{figure}[ht!]
	\centering
	\includegraphics[width=0.8\linewidth, keepaspectratio]{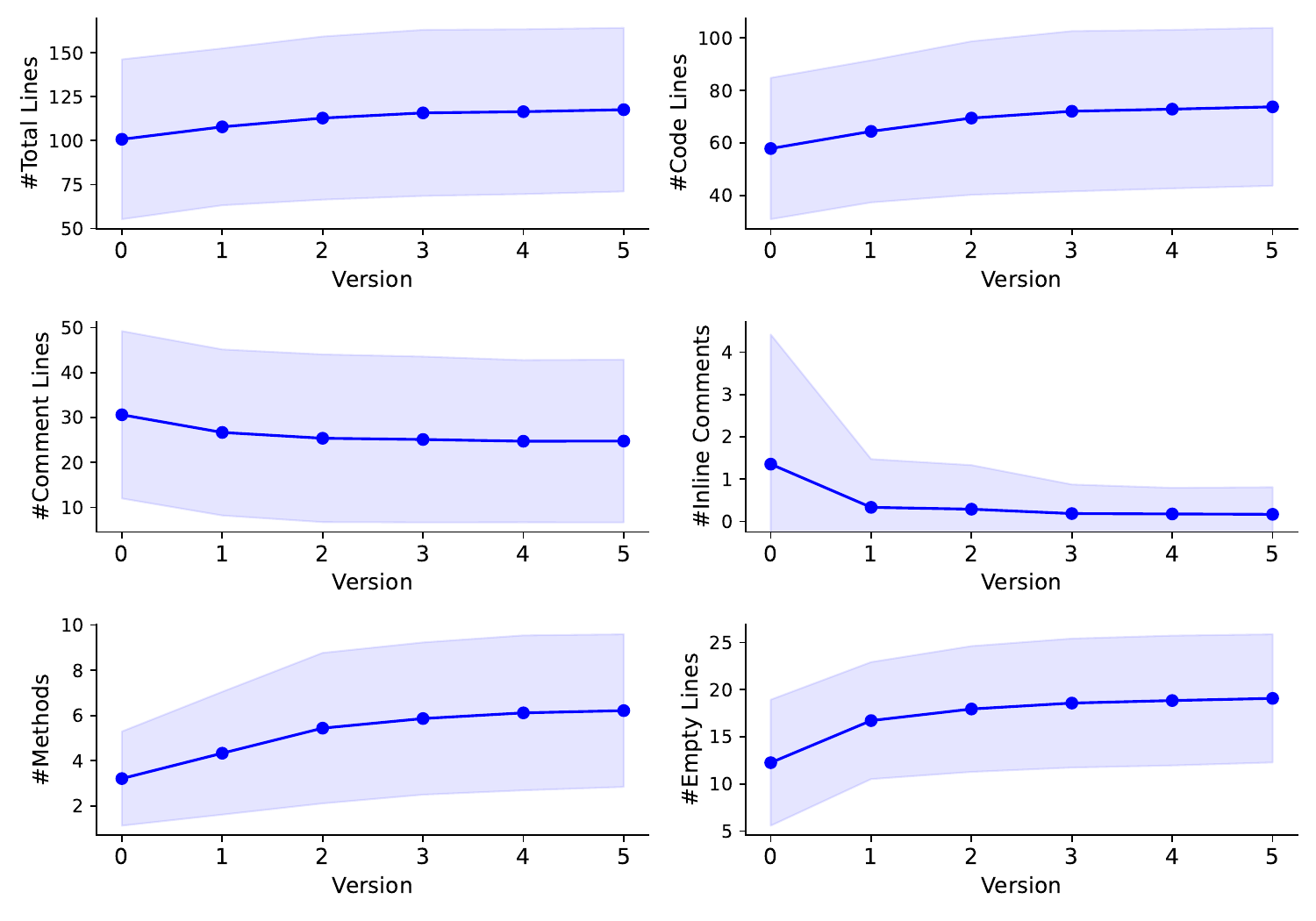}
	\caption{Average absolute metrics across refactorings (\VarOriginal, Prompt\textsubscript{General})). The shaded areas indicate standard deviation.}
    \Description{TODO}
	\label{fig:absolute_values_across_all-snippets_KF0_only}
\end{figure}

GPT5.1 initially substantially refactors code with fewer changes in each subsequent iteration. We visualize the change in average structural properties of all code snippets for \VarOriginal~based on Prompt\textsubscript{General} in 
Figure~\ref{fig:absolute_values_across_all-snippets_KF0_only}.

During the first iteration (v0~$\rightarrow$~v1), an increase in total lines can be observed, driven by a sharp increase in empty lines but also an increase in code lines. In contrast, the average number of comment lines decreases slowly, while inline comments are almost completely removed.
Meanwhile, the number of code lines increases steadily from version v0 to v5, rising from 58 to over 73. A similar upward trend is evident in the number of empty lines and number of method declarations, both of which similarly stabilize from version v3 onward.

\subsubsection{Overall Change Dynamics Across Refactoring Iterations (\VarOriginal)}

\begin{figure}[ht!]
	\centering
	\includegraphics[width=0.85\linewidth, keepaspectratio]{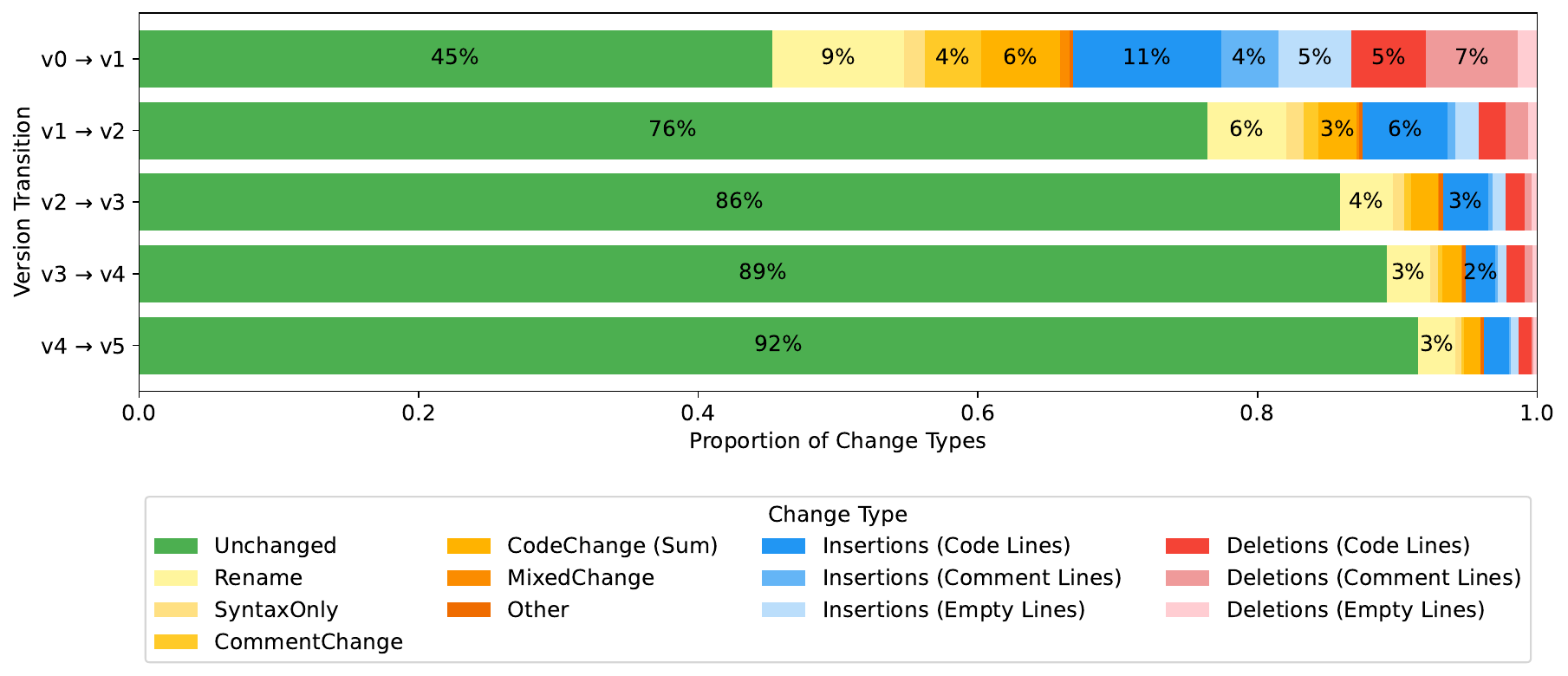}
	\caption{Detailed proportional distribution of code changes across iterations (\VarOriginal, Prompt\textsubscript{General}).}
    \Description{TODO}
	\label{fig:stacked_bar_plot_horizontal_main-exp_KF0_pKF0_detailed}
\end{figure}

During the first iteration (v0~$\rightarrow$~v1), 45\% of the lines remain unchanged. In the following iteration (v1~$\rightarrow$~v2), the proportion of unchanged lines increases to 76\%, and all types of changes decrease accordingly.
From v2~$\rightarrow$~v3 onward, the proportion of unchanged lines further increases, reaching 86\%, 89\%, and 92\% in the final iterations, respectively. Overall, the data indicate a consistent decrease in the extent of changes across refactoring iterations. 

To further examine the nature of the observed changes, we studied a finer-grained breakdown of the change types for \VarOriginal, which we visualize in Figure~\ref{fig:stacked_bar_plot_horizontal_main-exp_KF0_pKF0_detailed}.
Here, we differentiate the change types mentioned above into further subtypes, such as specific change types as well as line-level insertions and deletions.
In the first refactoring iteration (v0~$\rightarrow$~v1), the largest individual change types include renaming (9\%) and code insertions (11\%).
From v1~$\rightarrow$~v2, the changes become more targeted and balanced as changes are still observed but at lower frequencies.
In later iterations (v2~$\rightarrow$~v3 through v4~$\rightarrow$~v5), most change types fall below the 1\% threshold. The vast majority of lines remain unchanged, and no single change type dominates, suggesting a general stabilization of the refactoring.

\paragraph{Flow of Code Changes Across Iterations (\VarOriginal)}

\begin{figure}[ht!]
    \centering
    \includegraphics[width=0.6\linewidth, keepaspectratio]{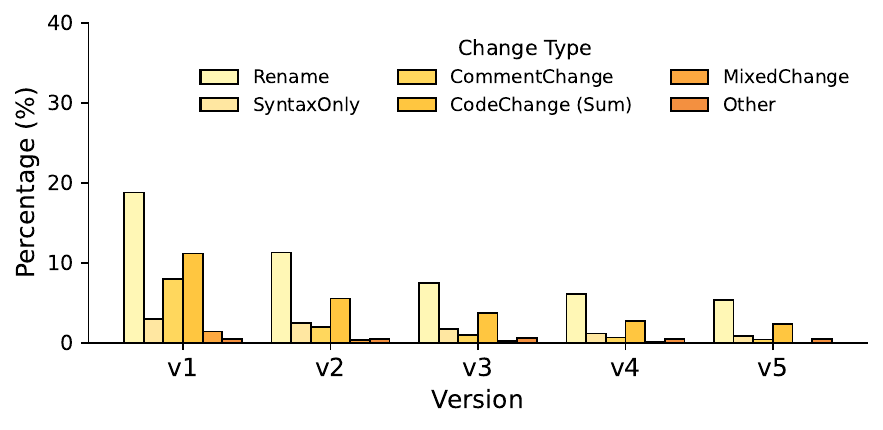}
    \caption{Percentage of different change types across iterations (\VarOriginal, Prompt\textsubscript{General}).}
    \Description{TODO}
    \label{fig:histplot_main-exp_KF0_pKF0}
\end{figure}

To better illustrate the progression and persistence of individual change types over the course of the refactoring, Figure~\ref{fig:histplot_main-exp_KF0_pKF0} presents the percentage of different change types for \VarOriginal~under Prompt\textsubscript{General}. While the stacked bar plot (Figure~\ref{fig:stacked_bar_plot_horizontal_main-exp_KF0_pKF0_detailed}) shows only aggregated proportions per version pair, this plot enables a more granular view of how specific change types evolve across all five iterations.

Rename operations and semantic changes dominate the early refactoring stages and persist across multiple iterations, while comment changes, syntax-only changes, and mixed changes occur less frequently and are more evenly distributed. Overall, the visualization highlights that even though the proportion of changes becomes marginal in later iterations, a diverse set of change types remains active throughout the refactoring.

\subsubsection{Pairwise Similarity Analysis Across Refactorings (\VarOriginal)}

\begin{figure}[ht!]
	\centering
	\includegraphics[width=0.5\linewidth, keepaspectratio]{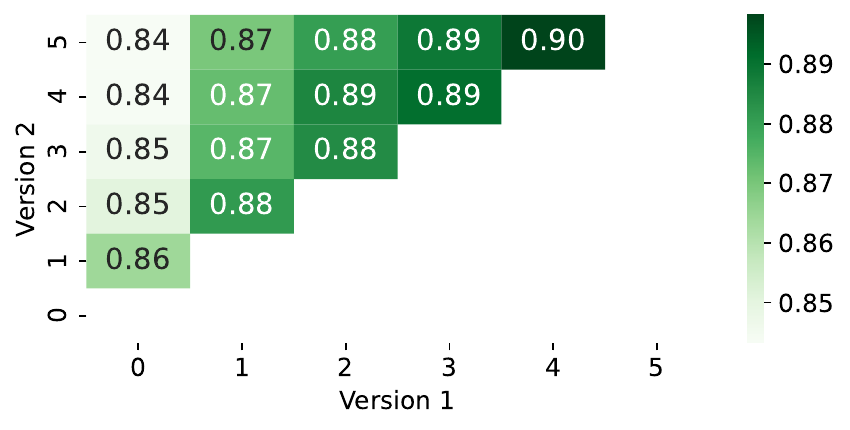}
	\caption{Pairwise similarity heatmap across refactorings (\VarOriginal, Prompt\textsubscript{General}). Each cell indicates the average similarity of changed lines between the respective version pair, using the metric described in Section~\ref{subsec:metrics}. Only the upper triangle is populated, as the comparison is directional (from earlier to later versions).}
    \Description{TODO}
	\label{fig:avg_simscore_heatmap_KF0_pKF0}
\end{figure}

While the previous analysis on provides a detailed view of how specific change types propagate across refactoring iterations, they do not reveal whether these changes lead to consistent structural convergence or occasional reversals. To address this aspect, we complement the change-type analysis with pairwise similarity scores between all versions. 
We visualize these scores in a heatmap allowing us to uncover convergence trends across refactorings and to identify potential back-and-forth changes. 
In Figure~\ref{fig:avg_simscore_heatmap_KF0_pKF0}, we show a heatmap of average similarity scores between all pairwise combinations of versions for \VarOriginal, based on Prompt\textsubscript{General}. 
It shows a steady increase in similarity as the refactoring progresses. The similarity between adjacent versions increases from 0.86 (v0~$\rightarrow$~v1) to 0.90 (v4~$\rightarrow$~v5). 
Non-consecutive comparisons (e.g., v0~$\rightarrow$~v5 = 0.84) also show an increased similarity, indicating that changes accumulate incrementally and consistently over iterations.

Notably, the similarity between early and late versions (e.g., v1~$\rightarrow$~v4 = 0.87, v2~$\rightarrow$~v5 = 0.88) suggests that structural convergence begins early in the refactoring and stabilizes in later iterations. However, none of the pairwise comparisons ever reach a perfect similarity score of 1.00.

\RQAnswer{RQ\textsubscript{1}}{
The results show that the iterative refactorings of already well-structured code (\VarOriginal) largely preserve the original structure and exhibit a clear convergence trajectory. However, the LLM did not strictly limit itself to only necessary changes. Initial iterations introduced a noticeable restructuring phase characterized by reductions in comments, renaming operations, and additional code adjustments. Subsequent refactorings stabilized quickly, though, with steadily increasing similarity scores and only marginal changes. Notably, the absence of perfect similarity values (1.00) indicates that LLMs continue to apply small changes even when the input already conforms to best practices, suggesting a potential tendency toward over-refactoring.}

\subsection{Convergence Across Code Variants (RQ\textsubscript{2})}
\label{sec:RQ2}
	
To address RQ\textsubscript{2}, we expand the analysis beyond the well-structured baseline (\VarOriginal) and investigate how LLMs handle systematically altered input. This setup with variants that violate typical readability conventions allows us to test whether the refactoring is sensitive to different starting conditions or whether it ultimately normalizes distinct variants toward a common structural representation. For this purpose, we generate two additional variants from each original snippet: one where we remove meaning from all identifiers and comments (\VarMeaningless~) and one where we remove all comments (\VarNoComment). All three variants (\VarOriginal, \VarMeaningless, \VarNoComment) are then subjected to five refactoring iterations using the same base Prompt\textsubscript{General} (\textit{"Refactor this code for improved readability."}). This design enables us to disentangle the effect of initial structural differences from the general convergence dynamics of iterative refactorings.

\subsubsection{Baseline: Variant Creation and Differences}

\begin{figure}[ht!]
	\centering
	\includegraphics[width=0.85\linewidth, keepaspectratio]{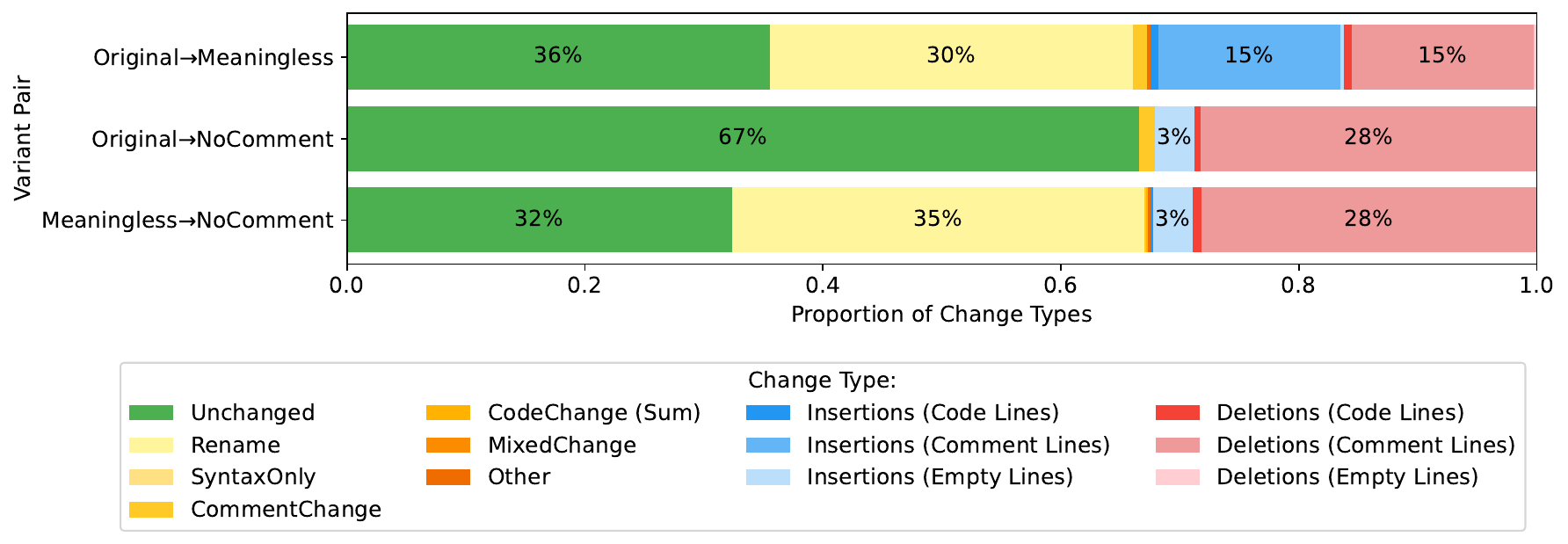}
	\caption{Detailed change composition between all three variants at v0 (i.e., our modifications before sending it to the LLM).}
    \Description{TODO}
	\label{fig:v0_detailed_average_changes}
\end{figure}

Before investigating the convergence of iteratively refactored code variants, we first examine how distinct the variants in their initial version v0 are. In Figure~\ref{fig:v0_detailed_average_changes}, we show the types of changes we applied between the three code variants (\VarOriginal, \VarMeaningless, \VarNoComment) in version v0 (i.e., before sending it to the LLM).

The modification \VarOriginal~$\rightarrow$ \VarMeaningless~consists of systematically renaming all class, method, and variable names, and changing the content of all comments to be meaningless. As expected, this results in a dominant proportion of rename operations (30\%), alongside deletions (15\%) and insertions of comments (15\%). The difference between these two change types arises from the matching algorithm: When a comment line is changed completely (rather than just an individual word), the changes are classified as an insertion and deletion (rather than a comment change).

In the case of \VarOriginal~$\rightarrow$ \VarNoComment, all code comments were removed, while keeping the functional code intact. On average, comments made up approximately 28\% of each snippet, which now appear as \textit{Deletions (Comment Lines)} in the comparison. Furthermore, about 3\% of all lines were replaced with empty lines to preserve code structure and maintain visual alignment. This choice was made deliberately, as we consider the insertion of empty lines to be a more appropriate structural substitute for removed comments, especially in terms of readability and comparability.

The third comparison (\VarMeaningless~$\rightarrow$ \VarNoComment) reflects the cumulative structural impact of both readability aspects: renaming and comment removal. On average, 68\% of all lines are affected: 35\% by renaming, 28\% by comment deletions, and 3\% are empty line insertions. Only 32\% of the lines remain unchanged. Note that no additional \textit{CommentChange} entries could be recorded in this comparison, as the \VarNoComment~contains no comments, thereby eliminating the possibility of a match in this change type.
Across all comparisons, only negligible residuals of other change types ($\leq$1\%) appear, which reflect inherent limitations of the parsing process.

\subsubsection{Absolute Code Metrics Across Iterations (\VarMeaningless~+ \VarNoComment)}

\begin{figure}[ht!]
	\centering
	\includegraphics[width=0.8\linewidth, keepaspectratio]{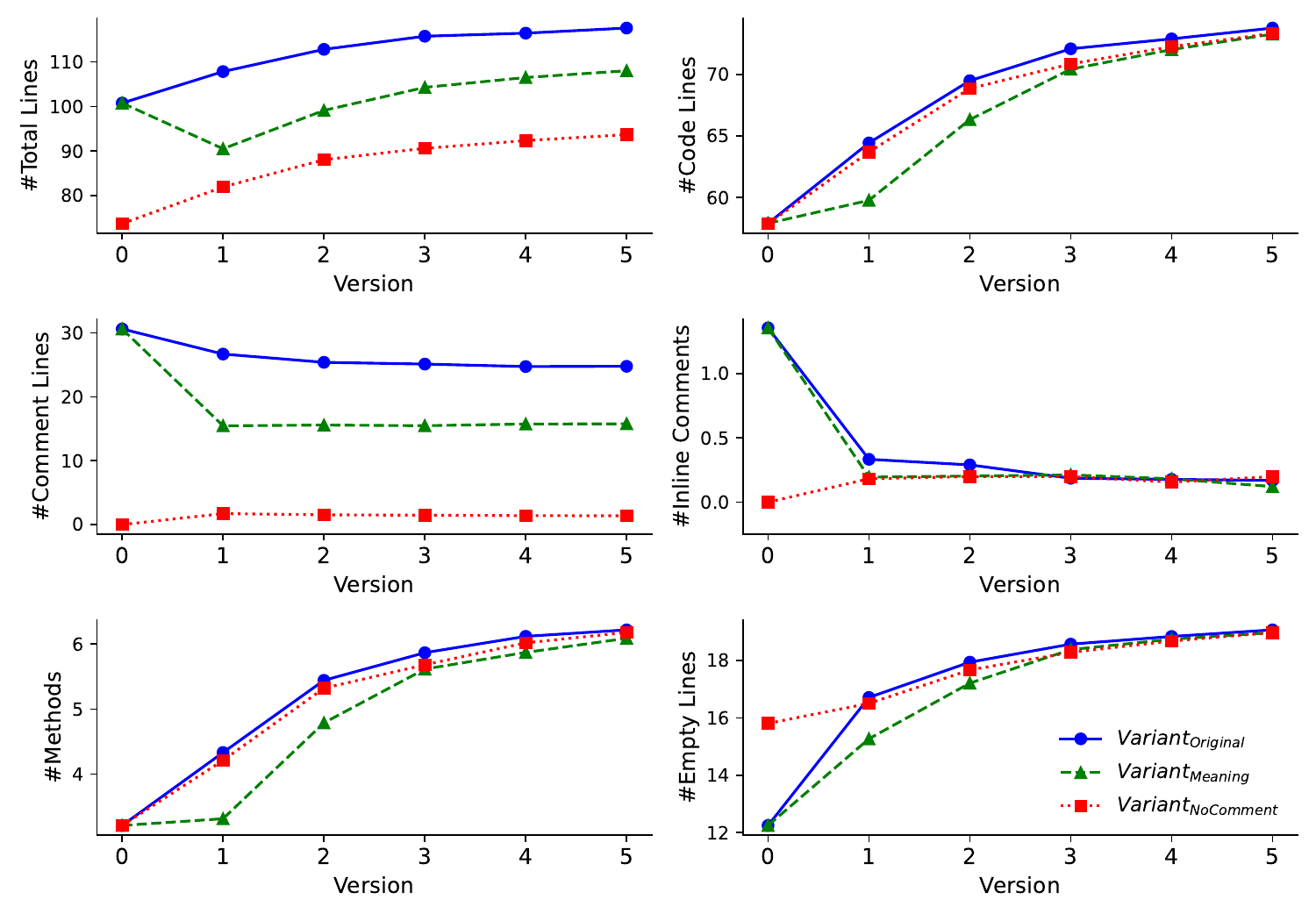}
	\caption{Average absolute metrics across refactoring iterations for \VarOriginal–\VarNoComment~(Prompt\textsubscript{General})}
    \Description{TODO}
	\label{fig:absolute_values_across_all-snippets_pKF0}
\end{figure}

First, we examine the structural changes across iterations, now also considering the additional variants. In Figure~\ref{fig:absolute_values_across_all-snippets_pKF0}, we visualize the evolution of absolute structural metrics across all refactoring iterations (v0 to v5) for the three code variants (\VarOriginal, \VarMeaningless, \VarNoComment), each refactored using general refactoring prompt. While the starting points at v0 are different due to our modifications, the changes are of similar nature, and for some metrics the values at the end are very similar. 

At version v0, \VarOriginal~and \VarMeaningless~contain 101 total lines on average, respectively, while \VarNoComment, due to the removal of all comments, begins with only 72 lines. After the first refactoring (v1), the LLM removed a lot of comments from \VarMeaningless, but otherwise the changes across all three variants are similar. There is an increase in code lines and empty lines. From version v2 onward, total line counts increase slightly and stabilize at different levels for each variant, almost exclusively due to the different number of comment lines.

The number of code lines increases steadily across all variants. All variants begin with 56 code lines and reach approximately 73 in v5.
Inline comments follow an interesting pattern: \VarOriginal~and \VarMeaningless~start at 1.3, and drop to 0.2 at v5. \VarNoComment~naturally starts with no inline comments, but converges to the very low number of roughly 0.2 at v5.
Empty lines steadily increase across all conditions and iterations. Lastly, the average number of methods also increases for all three variants, especially in the early iterations. All variants naturally start with the same number of methods (3.1) and despite different trajectories all end up at roughly 6 methods for each snippet, further indicating structural convergence.

\subsubsection{Overall Change Dynamics Across Refactoring Iterations (\VarMeaningless~+ \VarNoComment)}

\begin{figure}[ht!]
	\centering

    (a) \VarMeaningless~under Prompt\textsubscript{General}
    
    \includegraphics[width=0.85\linewidth, keepaspectratio, trim={0 4.2cm 0 0}, clip]{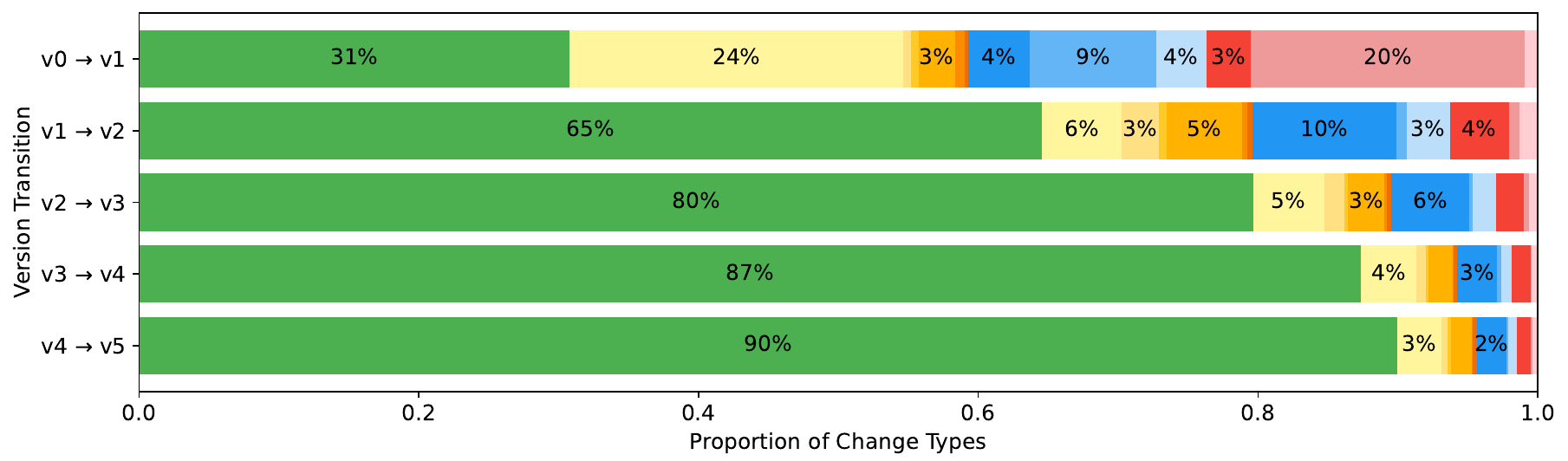}

    \vspace{0.5em}
    (b) \VarNoComment~under Prompt\textsubscript{General} \\
    
    \includegraphics[width=0.85\linewidth, keepaspectratio]{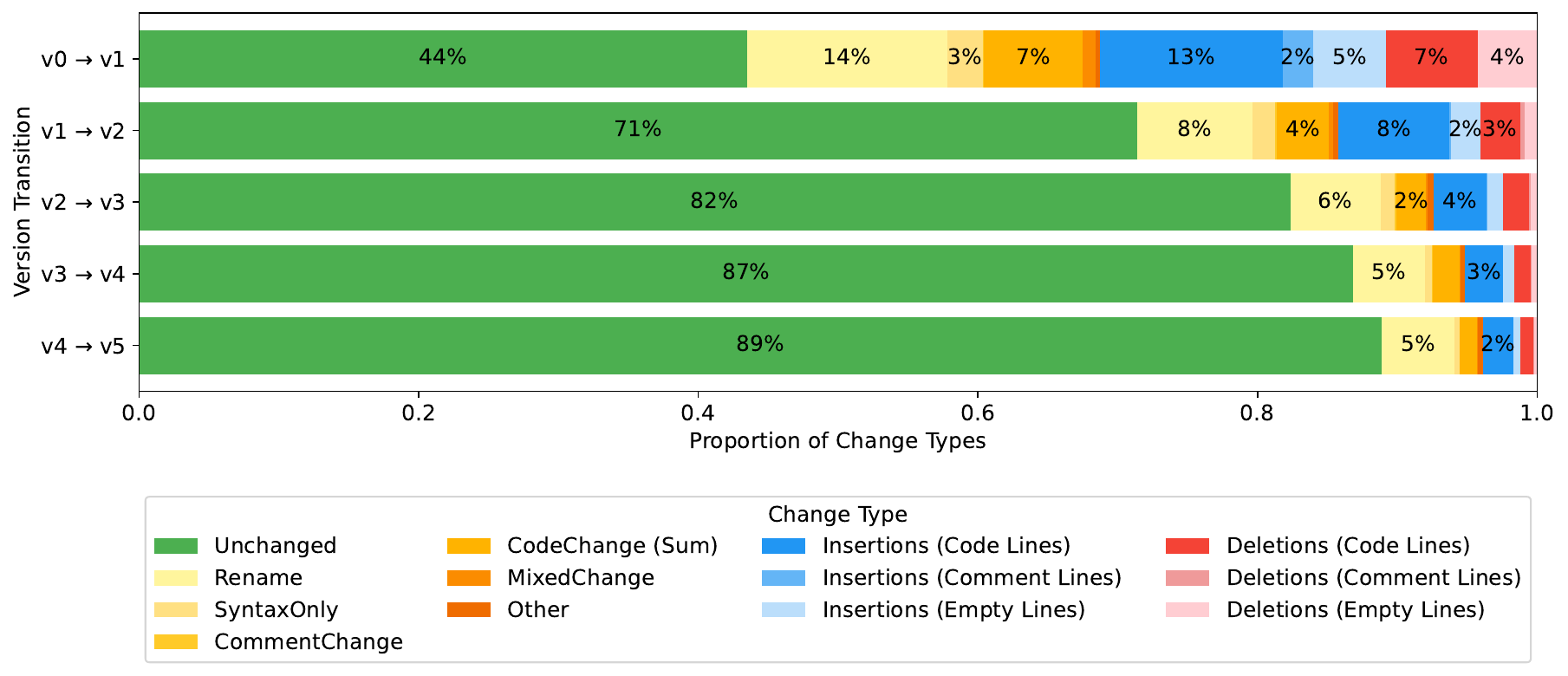}

	\caption{Comparison of average proportional change distributions across refactoring iterations for (a) \VarMeaningless~and (b) \VarNoComment, each under Prompt\textsubscript{General}.}
    \Description{TODO}
	\label{fig:stacked_bar_plot_horizontal_main-exp_KF1_KF2_pKF0_detailed}
\end{figure}

Having established the structural development of each variant in terms of absolute metrics, we now turn to a finer-grained view of how the underlying changes unfold across iterations. Previously, we captured overall trends in code size and composition. The following analysis focuses on the proportion and nature of line-level changes, that is how much of the code is changed, inserted, or deleted at each refactoring iteration. This perspective allows us to assess the degree and timing of stabilization during the iterative refactoring. 
We again visualize the average proportional distribution of changes across all refactoring iterations where we differentiate between unchanged lines, code changes, insertions, and deletions, providing a high-level view of how each version evolves over time. Figure~\ref{fig:stacked_bar_plot_horizontal_main-exp_KF1_KF2_pKF0_detailed} illustrates the data for \VarMeaningless~and \VarNoComment, both processed using the general refactoring prompt.

In the case of \VarMeaningless, the first refactoring iteration (v0~$\rightarrow$~v1) introduces substantial changes, with only 31\% of lines remaining unchanged. The following iterations show a clear trend toward stabilization: The proportion of unchanged lines increases to 65\%, 80\%, 87\%, and 90\%, respectively. Correspondingly, changes and structural operations (insertions/deletions) diminish steadily. The trend is similar for \VarNoComment, although the first iteration (v0~$\rightarrow$~v1) shows a slightly higher preservation rate, with 44\% of lines unchanged. In subsequent iterations, the proportion of unchanged lines increases to 71\%, 82\%, 87\%, and 89\%, respectively, closely matching the trajectory observed for \VarMeaningless.

Overall, both variants exhibit a convergence pattern similar to that of \VarOriginal~(cf.~Figure~\ref{fig:stacked_bar_plot_horizontal_main-exp_KF0_pKF0_detailed}), with the majority of structural adjustments occurring in the first two refactoring iterations. The final iterations incur only minimal change activity, suggesting that LLMs reach a stable representation of the code regardless of the starting condition.

\paragraph{Detailed Breakdown of Change Types (\VarMeaningless~+ \VarNoComment)}

While the aggregated values provide an overview of the general change dynamics, a finer-grained breakdown is required to understand the nature of the underlying changes. We break down each change type into the previously introduced specific change subtypes, such as renaming, syntax-only edits, or changes to comments, alongside different forms of insertions and deletions. This allows for a more nuanced assessment of how each refactoring iteration contributes to the convergence behavior observed earlier.

As expected, for \VarMeaningless~the largest change type in the first refactoring (v0~$\rightarrow$~v1) consists of rename operations (24\%). This is consistent with the intended purpose of \VarMeaningless, which targets identifier renaming and making comments meaningful. In the second iteration (v1~$\rightarrow$~v2), the rename portion drops to 6\%. From the third iteration onward, structural edits become increasingly rare: v2~$\rightarrow$~v3 and onward contain only minor changes across all change subtypes ($\leq$5\%), similar to the convergence trend observed in \VarOriginal.

A similar pattern appears for the \VarNoComment, albeit with a different emphasis. 
The initial refactoring (v0~$\rightarrow$~v1) shows 14\% of renames. The absence of comment changes and comment deletions reflect the absence of comments in the initial version v0 of the \VarNoComment. In subsequent iterations, the proportion of unchanged lines increases quickly and, eventually, reaches 89\%. From v2 onward, changes are smaller ($\leq$5\%) and mostly limited to minor renaming changes and more isolated insertions.

\paragraph{Flow of Changes Across Iterations (\VarMeaningless~+ \VarNoComment)}

To complement the proportional change distribution plots presented above, we again visualize the flow and persistence of change types across iterative refactoring. While stacked bar plots offer aggregated proportions per version pair, we visualize how individual types of changes propagate over time and to what extent they persist or fade. This is particularly relevant in the context of RQ\textsubscript{2}, where multiple code variants are initialized with targeted structural differences and may exhibit distinct convergence trajectories.

\begin{figure}[ht!]
    \begin{minipage}{0.48\textwidth}
        \centering
        \includegraphics[width=\textwidth, keepaspectratio]{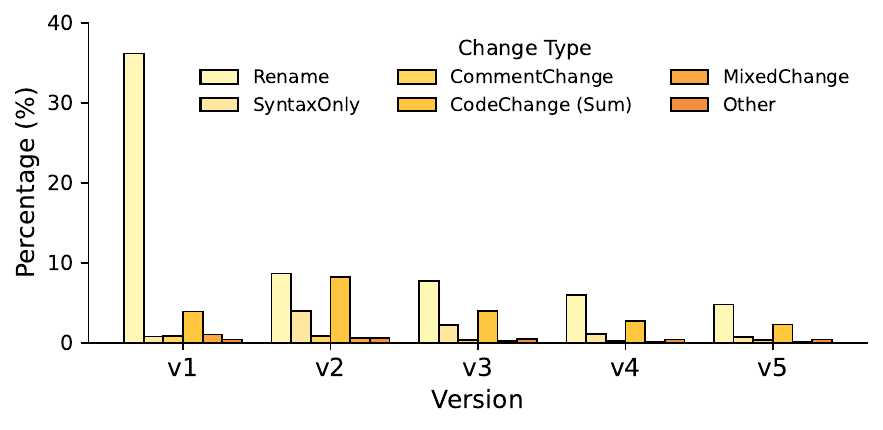}
        \caption{Percentage of change types across iterative refactorings for \VarMeaningless.}
        \Description{TODO}
        \label{fig:histplot_kf1_pKF0}
    \end{minipage}
    \hfill
    \begin{minipage}{0.48\textwidth}
        \centering
        \includegraphics[width=\textwidth, keepaspectratio]{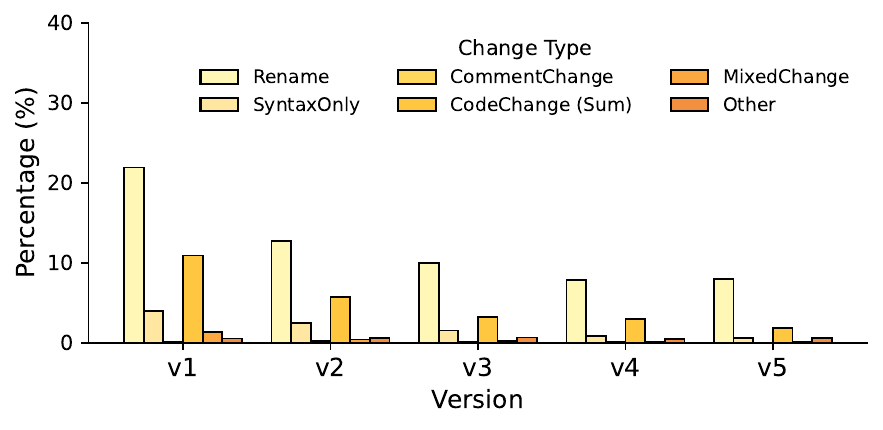}
    	\caption{Percentage of change types across iterative refactorings for \VarNoComment.}
        \Description{TODO}
	\label{fig:histplot_kf2_pKF0}
    \end{minipage}
\end{figure}

Figures~\ref{fig:histplot_kf1_pKF0} and \ref{fig:histplot_kf2_pKF0} present the relative changes and trajectories for \VarMeaningless~and \VarNoComment. In both cases, the initial iteration (v0 $\rightarrow$ v1) is dominated by the systematic renaming of identifiers. \VarNoComment~further exhibits an initial concentration of code changes and syntax-only changes.

Beyond the second iteration (v2 $\rightarrow$ v3 and onward), the distributions in both variants increasingly converge toward a narrow set of change types, with most activity limited to small-scale code changes, minor rename adjustments, and occasional syntax-only corrections. 
Overall, the change analysis reinforce the convergence trend already observed in the stacked bar plots and illustrate how different initial manipulations affect the temporal dynamics of refactoring.

\subsubsection{Pairwise Similarity Analysis Across Refactorings (\VarMeaningless~+ \VarNoComment)}

Following the structural and detailed analysis of code changes, we now examine how similar the generated code versions are to one another throughout the iterative refactoring. For this purpose, we analyze the average pairwise similarity scores between all versions from v0 to v5 for each variant. This measure complements the change-based analysis by capturing a more holistic view of code stability across iterations.

\begin{figure}[ht!]
    \begin{minipage}{0.48\textwidth}
        \centering
        \includegraphics[width=\textwidth, keepaspectratio]{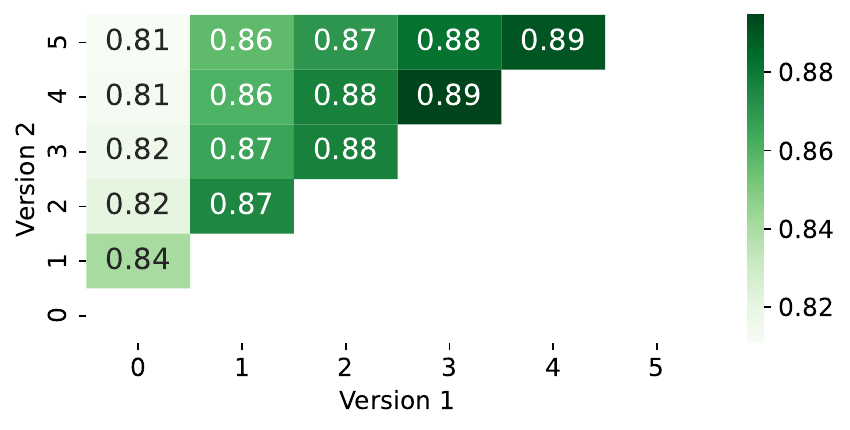}
        \caption{Similarity across iterations (\VarMeaningless, Prompt\textsubscript{General}).}
        \Description{TODO}
        \label{fig:avg_simscore_heatmap_KF1_pKF0}
    \end{minipage}
    \hfill
    \begin{minipage}{0.48\textwidth}
        \centering
    	\includegraphics[width=\textwidth, keepaspectratio]{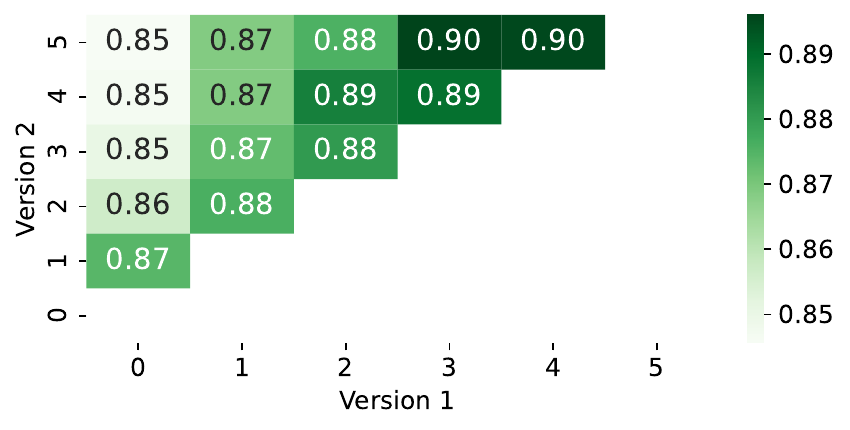}
    	\caption{Similarity across iterations (\VarNoComment, Prompt\textsubscript{General}).}
        \Description{TODO}
    	\label{fig:avg_simscore_heatmap_KF2_pKF0}
    \end{minipage}
\end{figure}

Figures~\ref{fig:avg_simscore_heatmap_KF1_pKF0} and \ref{fig:avg_simscore_heatmap_KF2_pKF0} presents the average similarity score heatmaps for the variants \VarMeaningless~and \VarNoComment, both processed under Prompt\textsubscript{General}. For both \VarMeaningless~and \VarNoComment, we observe a consistent pattern of increasing similarity in later iterations. However, the dynamics differ slightly between the two:
\VarMeaningless~shows a steady increase in similarity scores and suggest that refactorings stabilize significantly after the second iteration. 
This observation aligns with the earlier findings, which already indicated that most structural changes occur early, followed then by more incremental refactorings.
\VarNoComment, on the other hand, starts from a slightly higher similarity baseline, which reflects the more limited nature of the initial refactoring (comments) relative to \VarMeaningless~(renames of identifiers and making comments meaningful). Eventually, the similarity scores reach slightly higher values than in \VarMeaningless.

Overall, the iterative refactoring of both variants exhibit convergence behavior, with increasing similarity scores indicating that the iterative refactoring becomes more stable over time. Notably, \VarNoComment~converges minimally faster than \VarMeaningless, which aligns with the observation that its version transitions involve fewer structural alterations.

\paragraph{Similarity Score Evolution Across Variants}

\begin{figure}[ht!]
	\centering
	\includegraphics[width=0.85\textwidth]{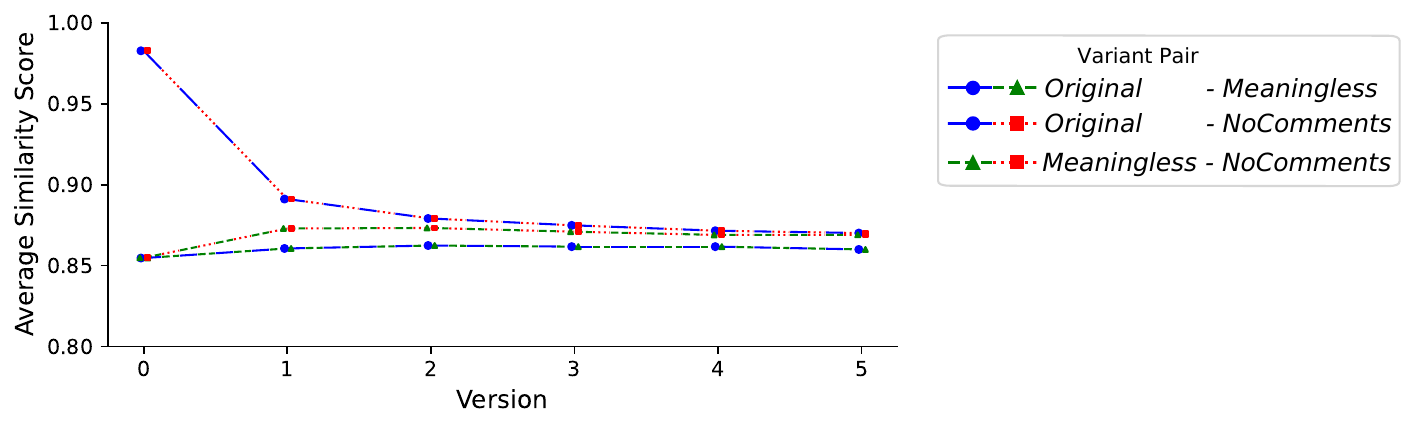}
	\caption{Development of average similarity scores across iterations for each variant pair under Prompt\textsubscript{General}.}
    \Description{TODO}
	\label{fig:lineplot_simscore_variant_comparison_pKF0}
\end{figure}

To complement the pairwise similarity heatmaps, Figure~\ref{fig:lineplot_simscore_variant_comparison_pKF0} illustrates how the similarity score between each variant pair evolves over the five refactoring iterations. This allows us to track whether and how the differences between variants persist or diminish throughout the refactoring.

At v0 (i.e., the starting point), the three variant pairs exhibit expected differences: First, \VarOriginal~and \VarNoComment~start out with the highest similarity score (0.98), due to both variants sharing identical code except for removed comments. \VarOriginal~and \VarMeaningless, and \VarMeaningless~and \VarNoComment, begin at a lower value (0.85), reflecting the impact of renamed identifiers and changed comments.

However, across iterations, all variant pairs gradually converge towards similar similarity scores ($\approx0.87$). Notably, the similarity between \VarOriginal~and \VarNoComment~decreases significantly from 0.98 to around 0.87, reflecting the increasing structural changes introduced by the LLM that go beyond simple formatting or syntax-only edits. The pair between \VarMeaningless~and \VarNoComment~exhibits a very slight upward trend from 0.85 to 0.87, closing in on the \VarOriginal--\VarMeaningless~score.

Ultimately, all three curves converge within a narrow range of less than 3\% difference, indicating that despite their divergent starting points, the iterative refactorings lead all variants to similar structural representations of the code, albeit still with some differences. In general, this provides further support for the convergence hypothesis formulated in RQ\textsubscript{2}: Independently of initial modifications, LLMs tend to normalize and align the structure of code snippets over repeated prompts.

\RQAnswer{RQ\textsubscript{2}}{
Our results show a general convergence across all three code variants. Despite structural differences in their initial versions (particularly between \VarMeaningless~and \VarNoComment), the iterative refactoring rapidly reduces these differences. 
Structural metrics (e.g., code, comment, and empty lines) largely stabilize after the second iteration, and line-level analyses reveal that the majority of changes occur early, followed by minimal adjustments in later iterations. 
Pairwise similarity scores confirm this trajectory: both within-variant comparisons (v0--v5) and cross-variant comparisons converge toward higher similarity values, typically around 0.87--0.90 by the final iteration. Notably, \VarNoComment~reaches convergence slightly faster than \VarMeaningless, but in the end all variants align to similar structural representations. 
Taken together, these findings suggest that LLM-driven refactoring is relatively robust to initial variations and tends to normalize code snippets toward a shared representation, thereby supporting the hypothesis of a convergence.}

\subsection{Impact of Explicitly Emphasizing Refactoring Aspects (RQ\textsubscript{3})}
\label{sec:RQ3}

To address RQ\textsubscript{3}, we conducted a comparative analysis of three prompt strategies designed to highlight distinct refactoring objectives. 
Specifically, in addition to our general prompt, we now ran the entire pipeline on all three variants again, but with Prompt\textsubscript{Meaning} explicitly directing LLMs toward meaningful identifiers, while Prompt\textsubscript{Comments} emphasizes the improvement of comments. 
By analyzing pairwise similarity scores across successive refactorings of the three code variants (\VarOriginal, \VarMeaningless, \VarNoComment), we aim to determine whether the explicit emphasis on a refactoring aspect leads to more stable convergence, or whether it introduces recurring back-and-forth changes. 

\subsubsection{Absolute Code Metrics Across Iterations Under Different Prompts}

\begin{figure}[ht!]
    \begin{minipage}{0.49\textwidth}
        \centering
        Prompt\textsubscript{General} vs. Prompt\textsubscript{Meaning}
        \includegraphics[width=\textwidth, keepaspectratio]{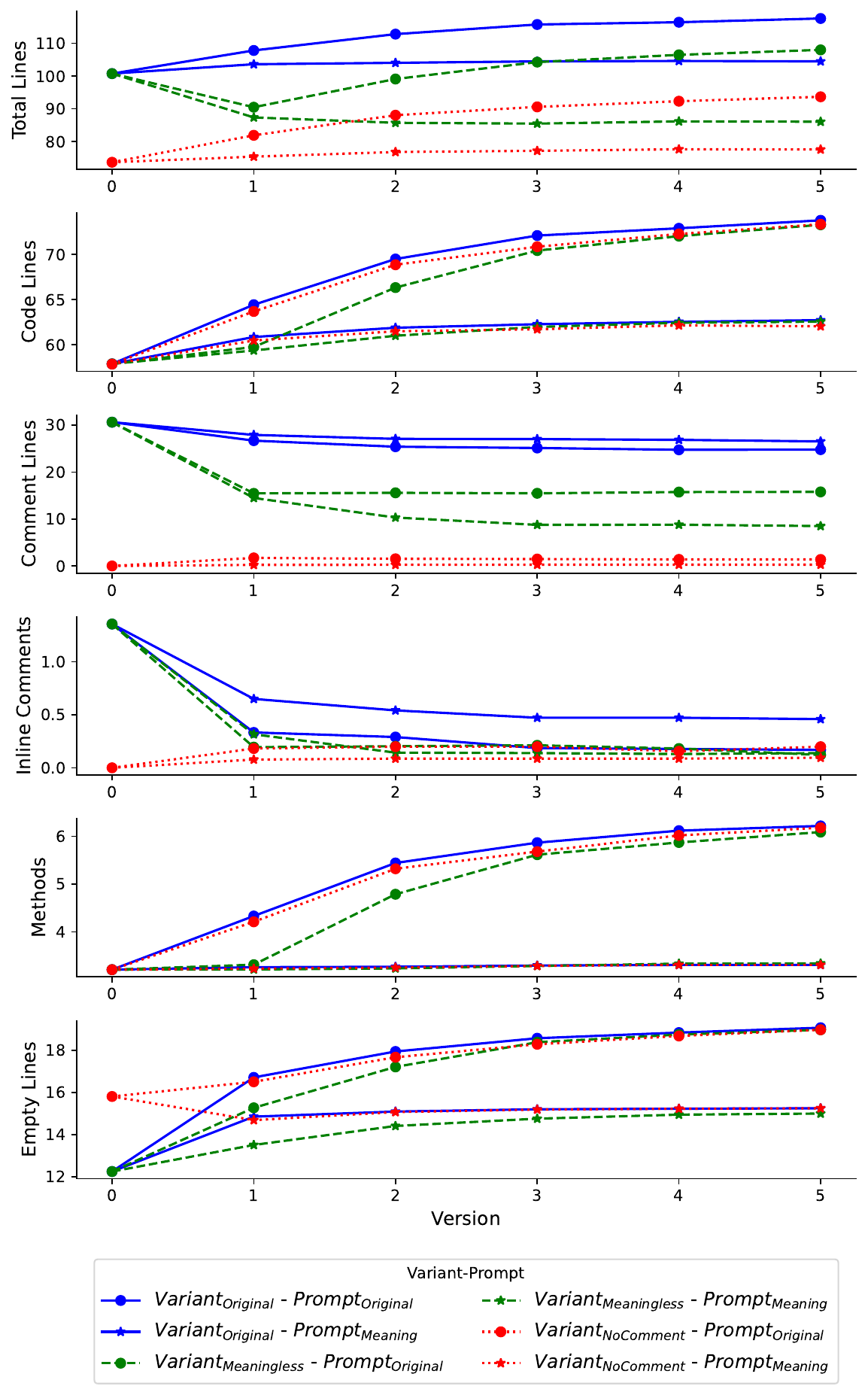}
        
        \caption{Comparison of absolute metrics across refactoring iterations for all variants for Prompt\textsubscript{General} and Prompt\textsubscript{Meaning}.}
        \Description{TODO}
        \label{fig:absolute_values_across_all-snippets_pKF1}
    \end{minipage}
    \hfill
    \begin{minipage}{0.49\textwidth}
        \centering
        Prompt\textsubscript{General} vs. Prompt\textsubscript{Comments}
    	\includegraphics[width=\textwidth, keepaspectratio]{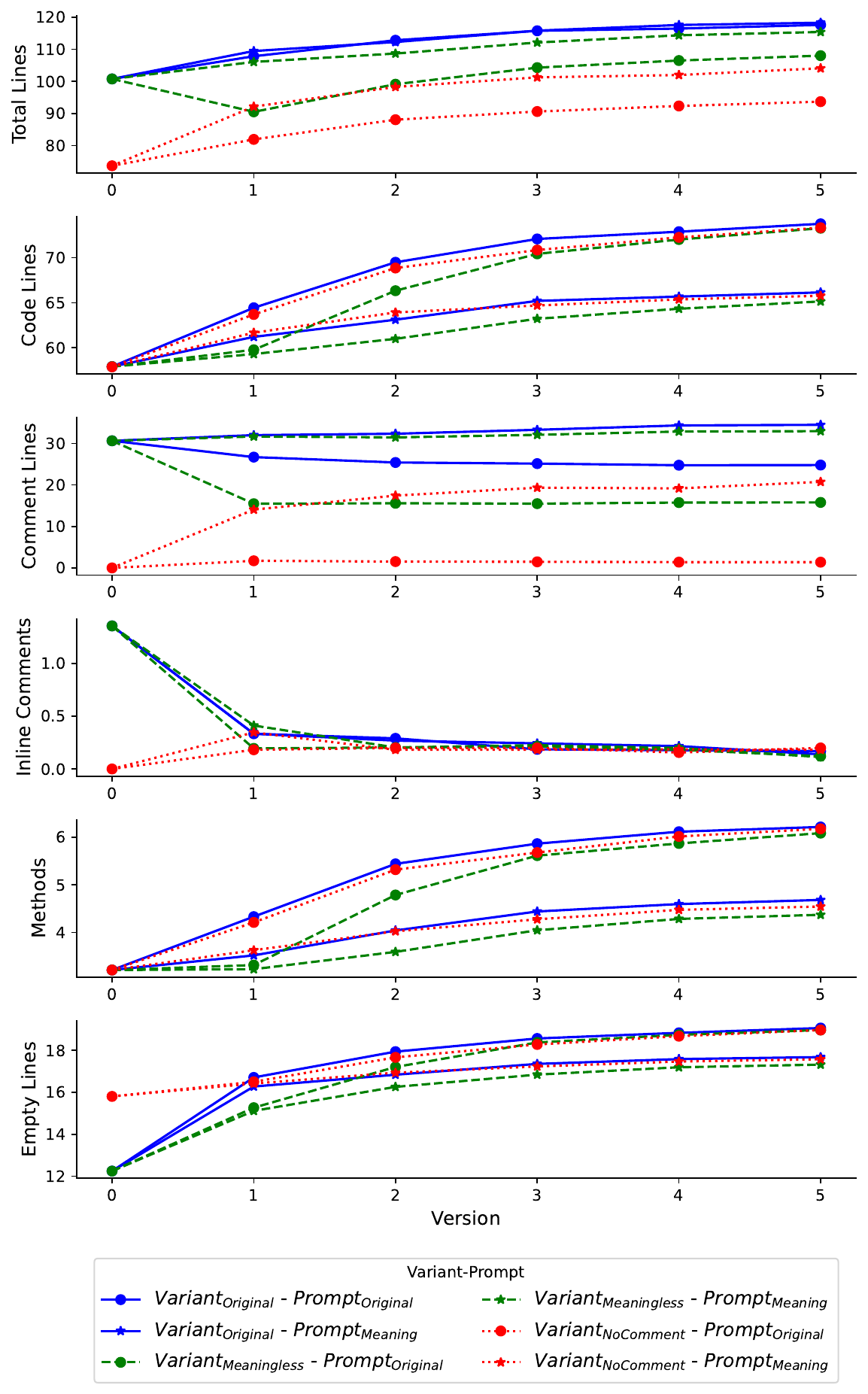}
        \caption{Comparison of absolute metrics across refactoring iterations for all variants for Prompt\textsubscript{General} and Prompt\textsubscript{Comments}.}
        \Description{TODO}
    	\label{fig:absolute_values_across_all-snippets_pKF2}
    \end{minipage}
\end{figure}

To investigate the effect of explicitly prompting for a particular refactoring aspect, we compared the absolute structural metrics of the snippets across all three variants under different prompt strategies. In Figures~\ref{fig:absolute_values_across_all-snippets_pKF1} and \ref{fig:absolute_values_across_all-snippets_pKF2}, it is apparent that explicitly instructing the LLM to focus on naming (Prompt\textsubscript{Meaning}) or to focus on comments (Prompt\textsubscript{Comments}) leads to noticeably different structural dynamics compared to Prompt\textsubscript{General}.

Specifically, for Prompt\textsubscript{Meaning}, we observe that structural metrics (e.g., total lines, number of methods, and code lines) remain relatively stable over the refactoring iterations across all three variants. This is in contrast to the behavior under Prompt\textsubscript{General}, where structural changes (e.g., increases in method count and code lines) are more pronounced, particularly in the first two refactoring iterations. There is also a notable trend that empty lines are not added as much when using Prompt\textsubscript{Meaning}. For the \VarNoComment, the structure remains particularly stable under Prompt\textsubscript{Meaning}. There is essentially no change in the number of methods and comment lines, which demonstrates that the meaning-focused prompt does not implicitly trigger the addition of documentation or structural embellishment.

Conversely, the dynamics are different when the prompt explicitly emphasizes comments (Prompt\textsubscript{Comments}). Here, the number of code lines and methods changes more similar to the general improvement prompt. As expected, the number of line and inline comments is generally higher than for the Prompt\textsubscript{Meaning}. However, the number of inline comments remain at a very low level. These trends indicate that the LLM responds directly to the prompt by inserting comments where appropriate, primarily during the first iteration. 
However, the lack of further increases in the number of comment lines or inline comments in later versions suggests that LLMs do not artificially inflate documentation density, despite an explicit prompt emphasizing comments.

\subsubsection{Overall Change Dynamics Across Refactoring Iterations}

To assess how targeted prompts affect the nature and stability of iterative code changes, we analyze both the proportional distribution of change types across refactoring iterations and the longitudinal flow of change types over time.
By analyzing the overall change dynamics between successive refactoring iterations, we can assess whether prompts that emphasize specific readability aspects (i.e., naming or commenting) lead to more targeted and stable improvements, or whether they merely shift the types of changes being applied. This perspective allows us to investigate whether such prompts accelerate convergence by reducing unnecessary structural changes and promoting more consistent refactoring behavior.
The following subsections therefore analyze how the distribution of code changes evolves across versions when using Prompt\textsubscript{Meaning} and Prompt\textsubscript{Comments}, respectively.


\paragraph{\VarOriginal:  Prompt\textsubscript{Meaning} vs. Prompt\textsubscript{Comments}}

\begin{figure}[ht!]
	\centering

    (a) \VarOriginal~under Prompt\textsubscript{Meaning} \\
    
    \includegraphics[width=0.85\linewidth, keepaspectratio, trim={0 4.2cm 0 0}, clip]{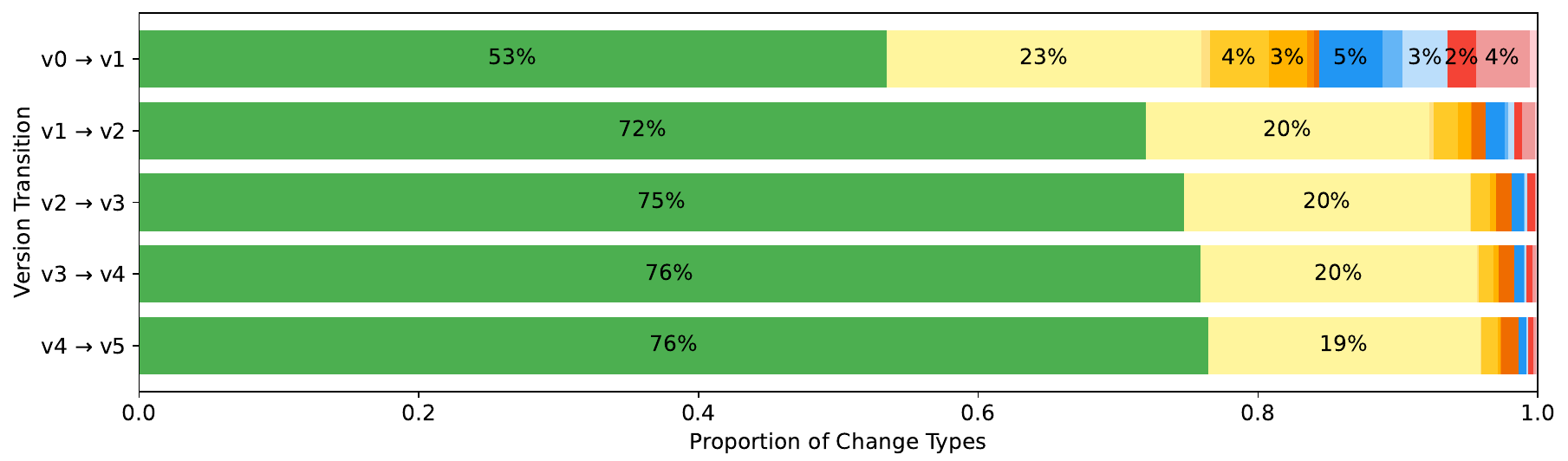}
    
    \vspace{0.5em}
    (b) \VarOriginal~under Prompt\textsubscript{Comments} \\
    
    \includegraphics[width=0.85\linewidth, keepaspectratio]{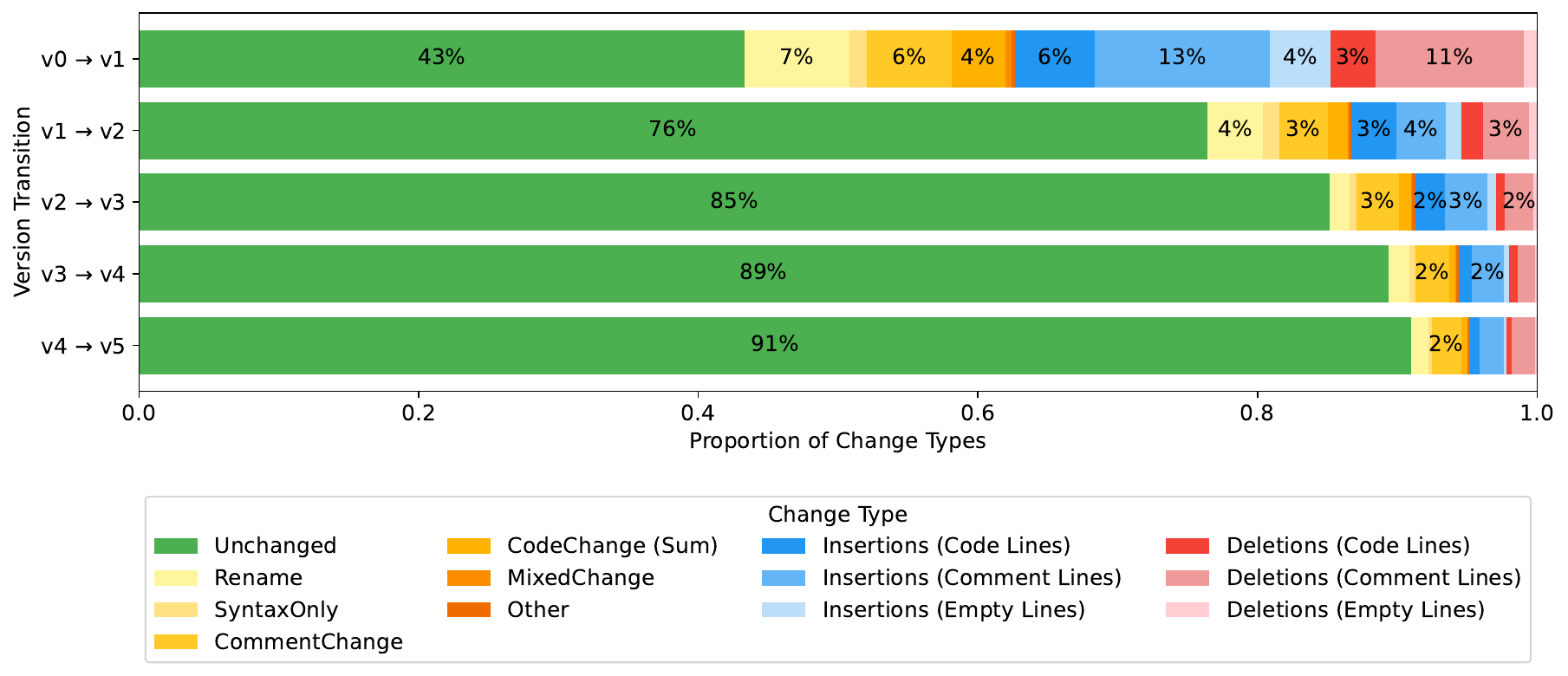}
    
	\caption{Distribution of change types for \VarOriginal~for (a) Prompt\textsubscript{Meaning} and (b) Prompt\textsubscript{Comments}.}
    \Description{TODO}
	\label{fig:stacked-KF0_pKF1_pFK2}
\end{figure}

In Figure~\ref{fig:stacked-KF0_pKF1_pFK2}, we present the detailed distribution of change types across successive refactoring iterations for \VarOriginal~under Prompt\textsubscript{Meaning} and Prompt\textsubscript{Comments}.
When we compare the two prompting strategies, we observe strikingly different trajectories of change types.
Under Prompt\textsubscript{Meaning}, rename operations dominate across all iterations and continue to a larger extent ($\approx20$\%) even in later iterations.
After the initial transition (v0~$\rightarrow$~v1), all changes besides renaming occur only marginally. 
In contrast, Prompt\textsubscript{Comments} produces a different pattern. The vast majority of changes concentrate in the first transition, where comment changes, insertions, and deletions  account for the bulk of activity.
After this initial refactoring, subsequent iterations show increasingly smaller changes.

\paragraph{\VarMeaningless: Prompt\textsubscript{Meaning} vs. Prompt\textsubscript{Comments}}

\begin{figure}[ht!]
	\centering

    (a) \VarMeaningless~under Prompt\textsubscript{Meaning} \\
    
    \includegraphics[width=0.85\linewidth, keepaspectratio, trim={0 4.2cm 0 0}, clip]{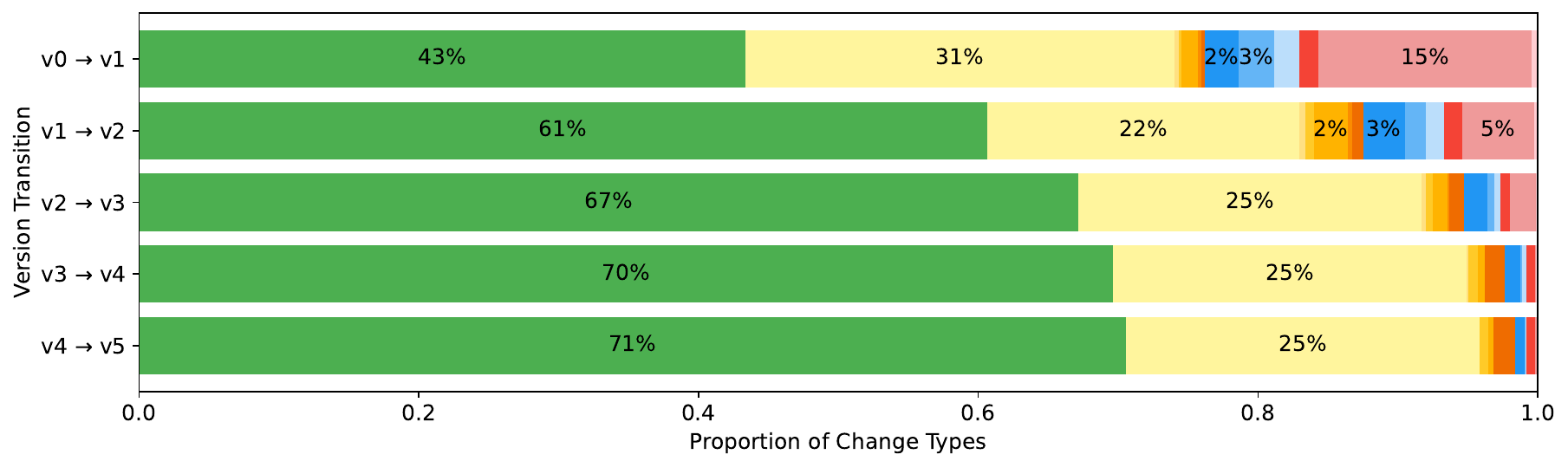}

    \vspace{0.5em}
    (b) \VarMeaningless~under Prompt\textsubscript{Comments} \\
    
    \includegraphics[width=0.85\linewidth, keepaspectratio]{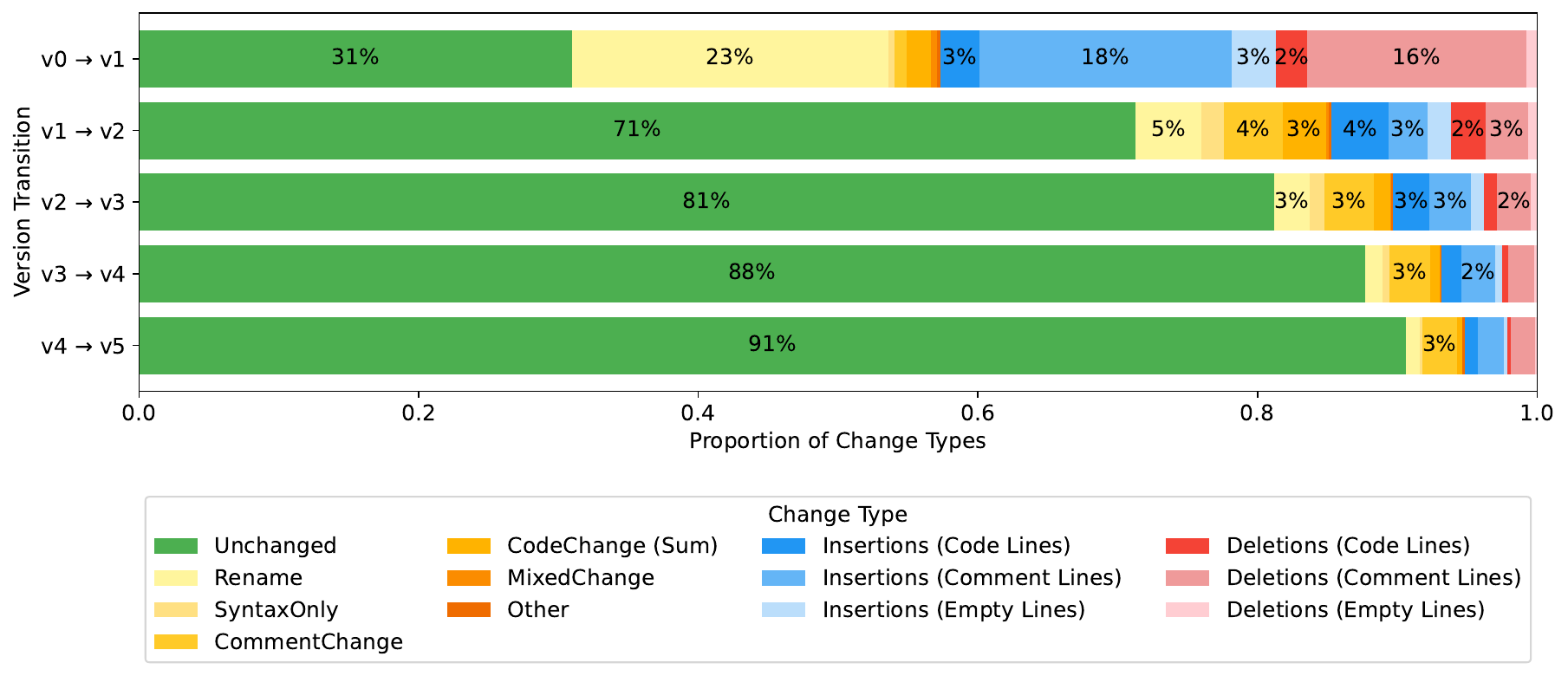}
    
	\caption{Distribution of change types for \VarMeaningless~for (a) Prompt\textsubscript{Meaning} and (b) Prompt\textsubscript{Comments}.}
    \Description{TODO}
	\label{fig:stacked-KF1_pKF1_pFK2}
\end{figure}

In Figure~\ref{fig:stacked-KF1_pKF1_pFK2}, we present the detailed distribution of change types across successive refactoring iterations for \VarMeaningless~under Prompt\textsubscript{Meaning} and Prompt\textsubscript{Comments}. Again, we find distinct effects of the two prompting strategies. Under Prompt\textsubscript{Meaning}, rename operations dominate from the very first iteration and persist across subsequent transitions, with an even increased activity pattern than observed in \VarOriginal. 
It is noteworthy that, unlike with the unguided Prompt\textsubscript{General}, the changes in \VarMeaningless~do not converge rapidly but instead show a persistent share of changes that decreases only slowly across iterations. 
Renaming remains the dominant change type throughout, while much smaller proportions of comment changes and other change types continue to appear in later iterations.

In contrast, Prompt\textsubscript{Comments} produces a more heterogeneous distribution of changes. Although the prompt emphasizes comments, the first transition (v0~$\rightarrow$~v1) still contains a large proportion of renames, likely reflecting the necessity of re-establishing meaningful identifiers.
In this initial refactoring, comment changes of all types account for around 36\% of all changes, demonstrating that LLMs generally follow the prompt. 
Similar to the Prompt\textsubscript{Meaning}, we continue to observe smaller changes in the code even in later iterations. However, changing comments remains the top change type until v5.

\paragraph{\VarNoComment: Prompt\textsubscript{Meaning} vs. Prompt\textsubscript{Comments}}

\begin{figure}[ht!]
	\centering

    (a) \VarNoComment~under Prompt\textsubscript{Meaning} \\
    
    \includegraphics[width=0.85\linewidth, keepaspectratio, trim={0 4.2cm 0 0}, clip]{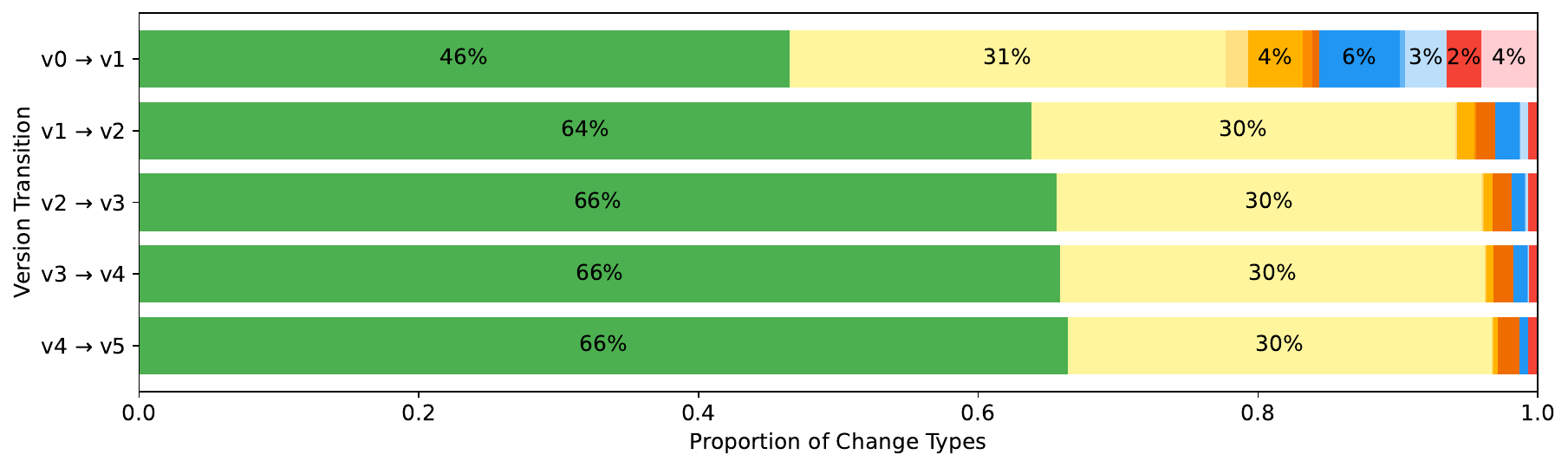}

    \vspace{0.5em}
    (b) \VarNoComment~under Prompt\textsubscript{Comments} \\
    
    \includegraphics[width=0.85\linewidth, keepaspectratio]{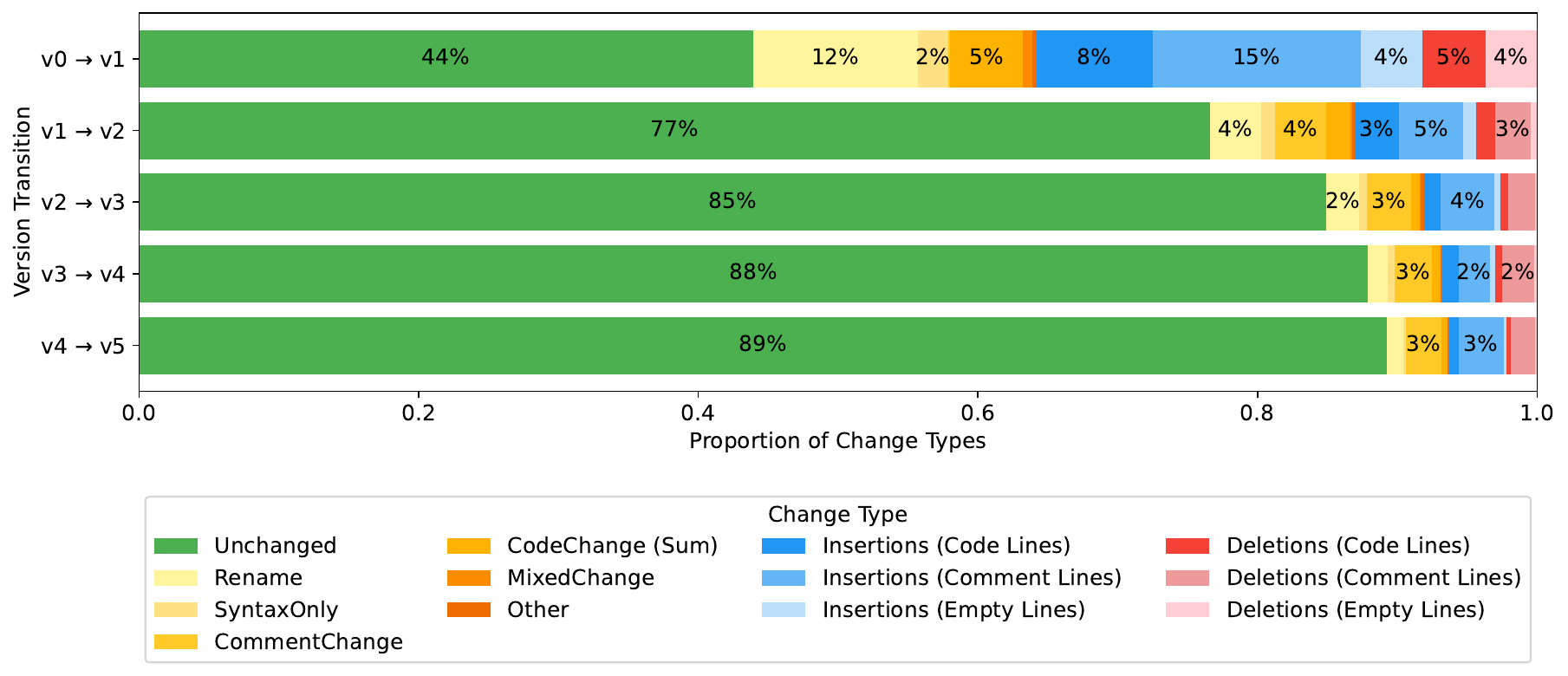}
    
	\caption{Distribution of change types for \VarNoComment~for (a) Prompt\textsubscript{Meaning} and (b) Prompt\textsubscript{Comments}.}
    \Description{TODO}
	\label{fig:stacked-KF2_pKF1_pFK2}
\end{figure}

In~Figure~\ref{fig:stacked-KF2_pKF1_pFK2}, we present the detailed distribution of change types across successive refactoring iterations for \VarNoComment~under Prompt\textsubscript{Meaning} and Prompt\textsubscript{Comments}. Here also, we find that the two prompting strategies drive markedly different change dynamics.
Under Prompt\textsubscript{Meaning}, rename changes dominate across all iterations, starting with a large proportion of renames in the initial transition (v0~$\rightarrow$~v1) and persisting through later iterations. 
Although the proportion of unchanged lines slightly increases with each iteration, renaming remains the most prominent activity throughout the refactoring. 

In contrast, Prompt\textsubscript{Comments} produces a pattern centered on comment changes. The first iteration shows a substantial share of comment changes, insertions, and deletions. After this initial iteration, activity rapidly diminishes, with subsequent transitions approaching near-stability and almost no further changes by the final iteration. 

\subsubsection{Pairwise Similarity Analysis Across Refactorings (Prompt\textsubscript{Meaning} vs. Prompt\textsubscript{Comments})}

\begin{figure}[ht!]
	\centering
	\includegraphics[width=\linewidth, keepaspectratio]{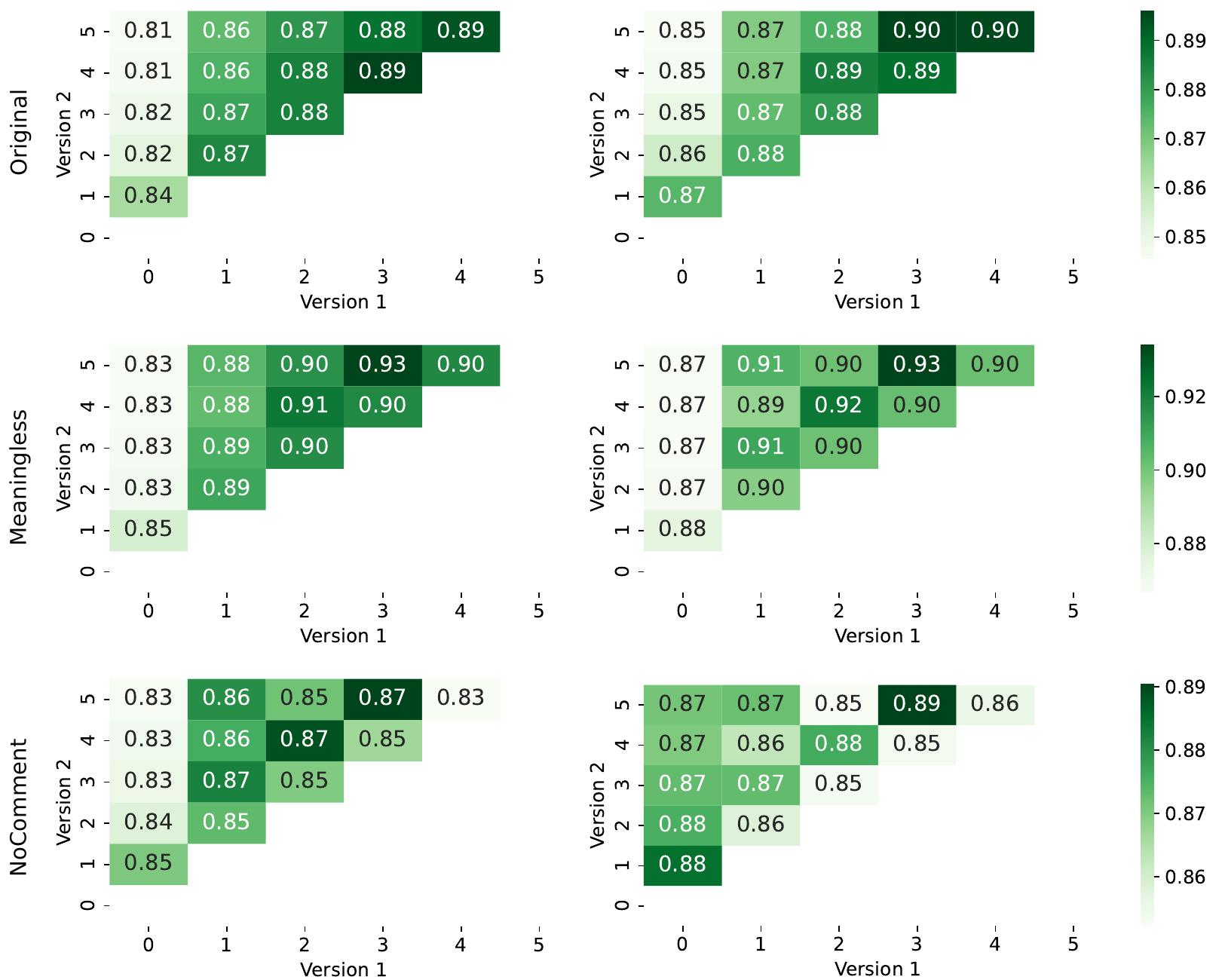}
	\caption{Similarity score heatmaps across all three code variants under Prompt\textsubscript{Meaning} (left) and Prompt\textsubscript{Comments} (right).}
    \Description{TODO}
	\label{fig:avg_simscore_heatmaps_all_variants_pKF1_pKF2}
\end{figure}

Next, we present the results of the pairwise similarity analysis across successive refactorings under the two prompt strategies, Prompt\textsubscript{Meaning} and Prompt\textsubscript{Comments}. 
We are interested in how similarity scores evolve across versions, and to identify whether different refactoring strategies lead to distinct convergence patterns or to recurring change behaviors. 
To illustrate these dynamics, we again consider heatmaps of the average similarity scores in Figure~\ref{fig:avg_simscore_heatmaps_all_variants_pKF1_pKF2}.

In the case of Prompt\textsubscript{Meaning}, we can observe different trends for each variant. While for the \VarOriginal~variant, there is a clear convergence with each iteration. To a large degree, this also holds true for the \VarMeaningless~variant, despite many changes. However, for \VarNoComment, the similarity scores are more chaotic, possibly due to the many alternating rename operations.

For Prompt\textsubscript{Comments}, there appear to be more back-and-forth changes, especially for the manipulated two variants. Starting from v1, alternating similarity scores become apparent, in which pairs of versions separated by two iterations share are very similar score, whereas the direct successor in between shows a lower score.

\paragraph{Similarity Score Evolution Across Variants (Prompt\textsubscript{Meaning} vs. Prompt\textsubscript{Comments})}

\begin{figure}[ht!]
	\centering
    \includegraphics[width=0.85\linewidth, keepaspectratio]{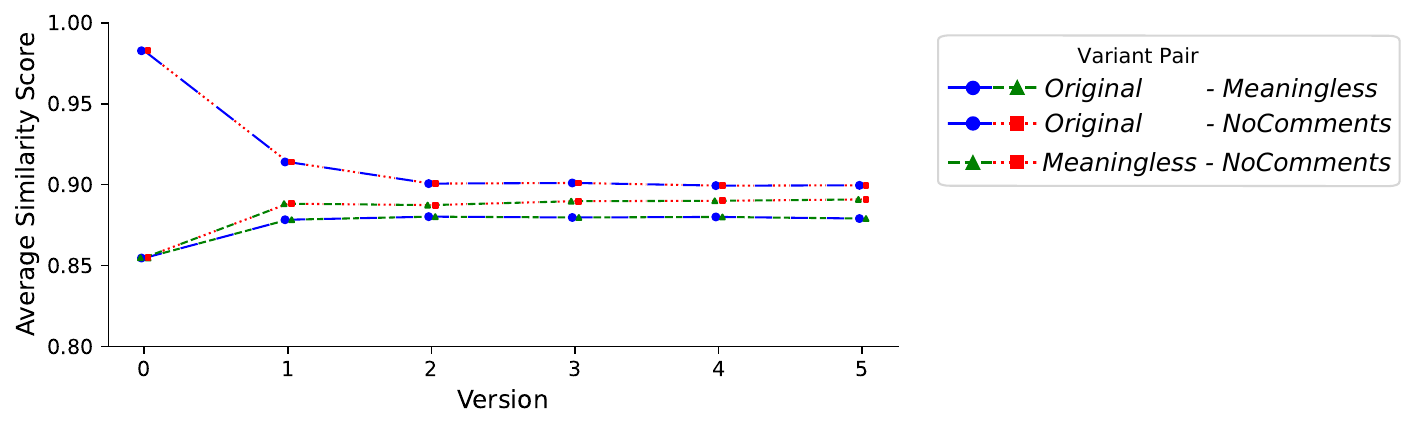}

    \includegraphics[width=0.85\linewidth, keepaspectratio]{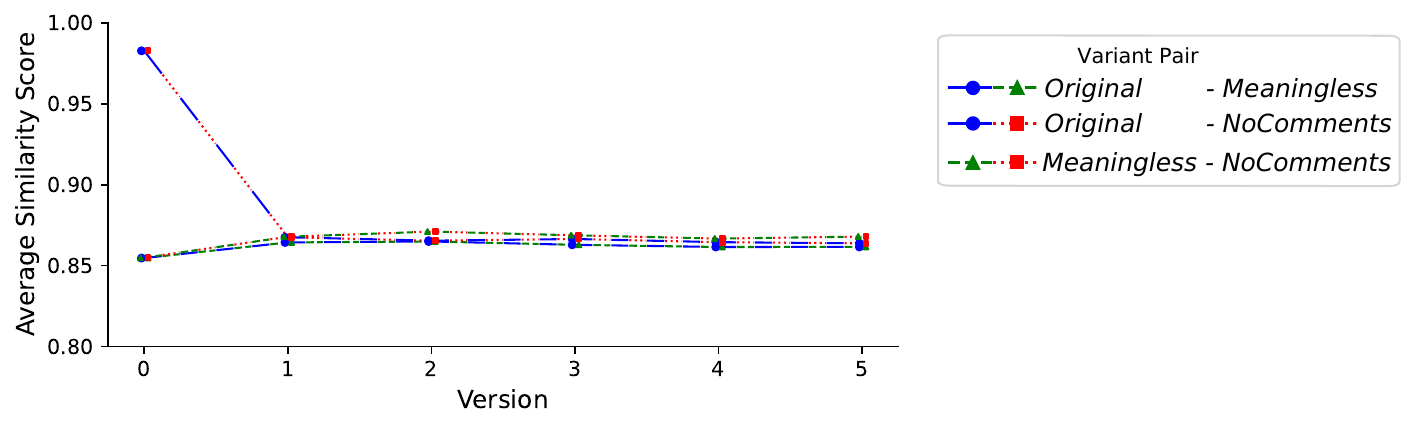}

	\caption{Convergence of similarity score across all code variants under Prompt\textsubscript{Meaning} (top) and Prompt\textsubscript{Comments} (bottom).}
    \Description{TODO}
	\label{fig:lineplot_simscore_variant_comparison_pKF1_pKF2}
\end{figure}

To directly compare the convergence tendencies across prompt strategies, we analyze the evolution of similarity scores across variant pairs. 
We illustrate in Figure~\ref{fig:lineplot_simscore_variant_comparison_pKF1_pKF2} that for both prompt strategies the similarity scores across the three variant pairs converge rapidly, with the most substantial change occurring during the transition from v0 to v1. Overall, Prompt\textsubscript{Meaning} achieves a higher average similarity score across the variant pairs compared to Prompt\textsubscript{Comments}.

\subsubsection{Statistical Analysis}

Building on the descriptive patterns observed so far, we applied statistical tests to determine whether the observed differences between prompts are statistically significant. Since our two targeted prompts focus on renaming of variables and change of comments, we focus our statistical analysis on the two appropriate metrics: \textit{Rename} and \textit{CommentChange}.

\paragraph{Descriptive Statistics}

\begin{table}[ht!]
	\centering
	\caption{Descriptive statistics (Mean $\pm$ SD) for the metric \textit{Rename} and \textit{CommentChange} for each variant and prompt.}
	\label{tab:descriptive-stats-rename-comment}
	\begin{tabular}{llrr}
		\toprule
		Code Variant & Prompt & Rename & CommentChange \\
		\midrule
		& Prompt\textsubscript{General} & 4.9\% $\pm$ 6.9\% & 1.2\% $\pm$ 2.7\% \\
		\VarOriginal~& Prompt\textsubscript{Meaning} & 20.5\% $\pm$ 9.3\% & 1.9\% $\pm$ 3.0\% \\
		& Prompt\textsubscript{Comments} & 3.1\% $\pm$ 5.4\% & 3.4\% $\pm$ 3.7\% \\
		\midrule
		& Prompt\textsubscript{General} & 8.3\% $\pm$ 10.3\% & 0.3\% $\pm$ 1.3\% \\
		\VarMeaningless~& Prompt\textsubscript{Meaning} & 25.6\% $\pm$ 10.4\% & 0.5\% $\pm$ 1.6\% \\
		& Prompt\textsubscript{Comments} & 6.4\% $\pm$ 9.5\% & 2.8\% $\pm$ 3.3\% \\
		\midrule
		& Prompt\textsubscript{General} & 7.9\% $\pm$ 9.8\% & 0.1\% $\pm$ 0.5\% \\
		\VarNoComment~& Prompt\textsubscript{Meaning} & 30.5\% $\pm$ 11.0\% & 0.0\% $\pm$ 0.5\% \\
		& Prompt\textsubscript{Comments} & 4.1\% $\pm$ 7.2\% & 2.5\% $\pm$ 3.6\% \\
		\bottomrule
	\end{tabular}
\end{table}

In Table~\ref{tab:descriptive-stats-rename-comment}, we report the mean percentages and standard deviations of \textit{Rename} and \textit{CommentChange} operations across all prompt variants and code versions. 
These descriptive statistics provide an intuitive understanding of the magnitude and variability of changes induced by different prompts. 
For instance, the renaming changes under Prompt\textsubscript{Meaning} consistently show higher means compared to Prompt\textsubscript{General} and Prompt\textsubscript{Comments}, such as in \VarNoComment~(30.5\% $\pm$ 11.0\%), suggesting a strong prompt-induced tendency toward renaming changes. 
Similarly, \textit{CommentChange} activity under Prompt\textsubscript{Comments} peaks in several configurations, indicating differential prompt sensitivity across change types. 
The high standard deviations also highlight the heterogeneity in the model's responses to different snippets.

\paragraph{Inferential Statistics with Kruskal-Wallis Test}

To evaluate the influence of different prompt variants on code refactoring, we applied the Kruskal-Wallis H test~\cite{Kruskal1952} as a non-parametric alternative to one-way ANOVA.
This test is particularly suited for our data, as the distributions of code change metrics are proportional (i.e., normalized between 0 and 1) and therefore not guaranteed to follow a normal distribution. 
We used the Kruskal-Wallis test to determine whether different prompts had a statistically significant effect on \textit{Rename} and \textit{CommentChange} within each of the three code variants. The test revealed significant effects of prompt variants on both \textit{Rename} and \textit{CommentChange} across all code variants:

\begin{itemize}
	\item Rename: Significant effects in \VarOriginal~($H = 1771.12$, $p < 0.001$), \VarMeaningless~($H = 1433.58$, $p < 0.001$), and \VarNoComment~($H = 1969.28$, $p < 0.001$).
	\item CommentChange: Significant effects in \VarOriginal~($H = 496.59$, $p < 0.001$), \VarMeaningless~($H = 1104.75$, $p < 0.001$), and \VarNoComment~($H = 1274.34$, $p < 0.001$).
\end{itemize}

\paragraph{Post-Hoc Pairwise Comparisons}

While the Kruskal-Wallis test indicates whether a difference exists among groups, it does not specify which groups differ from each other. 
To resolve this issue, we conducted post-hoc pairwise comparisons using the Mann-Whitney U test, which compares the distributions of two independent groups~\cite{Mann1947}. This allows us to identify which specific prompt pairs (e.g., Prompt\textsubscript{General} vs. Prompt\textsubscript{Meaning}) contributed to the overall significant differences.

\begin{table}[ht!]
	\centering
    \caption{Results of post-hoc tests with Mann-Whitney U test for the metrics \textit{Rename} and \textit{CommentChange} for each variant and prompt.}
	\label{tab:inferential-stats-rename-comment}
	\begin{tabular}{lllrr}
		\toprule
		Change Type & Code Variant & Prompt Comparison & Test Statistic & p-Value \\
		\midrule
		  & & General vs. Meaning & 123836 & $<0.001$ \\
        Rename & \VarOriginal & General vs. Comments & 784870 & $<0.001$ \\
        &  & Meaning vs. Comments & 1258480 & $<0.001$ \\
        \specialrule{0.1pt}{1pt}{1pt}
        & & General vs. Meaning & 172787 & $<0.001$ \\
        Rename & \VarMeaningless & General vs. Comments & 782883 & $<0.001$ \\
        & & Meaning vs. Comments & 1201370 & $<0.001$ \\
        \specialrule{0.1pt}{1pt}{1pt}
        & & General vs. Meaning & 100221 & $<0.001$ \\
        Rename & \VarNoComment & General vs. Comments & 846352 & $<0.001$ \\
        & & Meaning vs. Comments & 1288770 & $<0.001$ \\
        \midrule
        & & General vs. Meaning & 557780 & $<0.001$ \\
        CommentChange & \VarOriginal & General vs. Comments & 335010 & $<0.001$ \\
        & & Meaning vs. Comments & 449778 & $<0.001$ \\
        \specialrule{0.1pt}{1pt}{1pt}
        & & General vs. Meaning & 653612 & $0.176$ \\
        CommentChange & \VarMeaningless & General vs. Comments & 266403 & $<0.001$ \\
        & & Meaning vs. Comments & 291884 & $<0.001$ \\
        \specialrule{0.1pt}{1pt}{1pt}
        & & General vs. Meaning & 681923 & $<0.001$ \\
        CommentChange & \VarNoComment & General vs. Comments & 330944 & $<0.001$ \\
        & & Meaning vs. Comments & 319593 & $<0.001$ \\
		\bottomrule
	\end{tabular}
\end{table}

The post-hoc tests confirmed that these effects of prompt variants on the change types were driven by significant differences between all individual prompt conditions. Except for \textit{CommentChange} metric of the contrast between the Prompt\textsubscript{General} and Prompt\textsubscript{Meaning} within the \VarMeaningless, everything is highly statistically significant (i.e.,~$p<0.001$). 
These findings provide robust statistical evidence that the choice of prompt has a significant and consistent influence on how LLMs perform both renaming operations and comment changes, across all examined code variants. Prompt\textsubscript{Meaning} consistently elicited a higher frequency of renaming, while Prompt\textsubscript{Comments} exerted stronger effects on comment-related changes. These findings suggest that explicitly emphasizing a refactoring aspect in the prompt statistically significantly shapes the trajectory of code changes.

\RQAnswer{RQ\textsubscript{3}}{Our results demonstrate that prompt design can substantially shape refactoring trajectories. When the prompt targeted identifier renaming (Prompt\textsubscript{Meaning}), refactorings were dominated by recurring rename operations without reaching stable convergence. In contrast, prompts emphasizing comments (Prompt\textsubscript{Comments}) induced an increase in comment density during the first refactoring, followed by a stabilization and smaller subsequent changes. Our statistical analysis demonstrated that the choice of the prompt has a statistically significant effect on the suspected change metrics targeting renaming and comment operations.}

\section{Discussion}
\label{ch:discussion}

The open question identified in the gap analysis of Section~\ref{sec:gap-analysis} was how LLMs iteratively refactor code, whether this happens in a stable and reliable manner, and what influence different starting conditions and prompt strategies have. In this section, we discuss the findings of our experiment in terms of our research questions and how they fill this gap. It interprets the observed refactoring patterns, compares them to existing work, and highlights overarching themes that emerge across the analyses. In doing so, it outlines how the results contribute to a deeper understanding of iterative code changes by LLMs.

\subsection{Summary of Key Findings}

Overall, the results demonstrate that iterative refactorings of LLMs follow a consistent trajectory across different conditions. For well-structured code (RQ\textsubscript{1}), LLMs largely preserved the original organization but nonetheless introduced unnecessary adjustments in early iterations before stabilizing toward a near-converged state. When applied to systematically varied code snippets (RQ\textsubscript{2}), the refactorings converged across all variants, indicating that LLMs normalize divergent inputs towards structurally similar representations over time. Finally, we found that explicit prompt design (RQ\textsubscript{3}) exerts a strong influence on refactoring trajectories: While prompts emphasizing identifier renaming induced never-ending changes, prompts highlighting comments facilitated stabilization. Taken together, these findings provide robust evidence that iterative refactorings by LLMs are characterized by early restructuring followed by a general convergence, with prompt design serving as a decisive factor in determining whether the iterative refactorings stabilize or remain more oscillatory.


\subsection{Interpretation for Evolution of Iterative Refactoring (RQ\textsubscript{1})}

The analysis of RQ\textsubscript{1} reveals that LLMs tend to change code even when it already adheres to established best practices. However, these changes stabilize after a couple of refactoring iterations. The most prominent structural refactoring involves encapsulating functionality into smaller functions, which on the one hand partially supports earlier findings that (iteratively) generated code often suffers from unnecessary or increasing complexity~\cite{liu_refining_2023, stein_code_2025}. On the other hand, this suggests that LLMs initially identify a substantial number of aspects to ``improve'', particularly when dealing with code that has not been previously refactored by an LLM. After only a few iterations, though, structural refactorings quickly stabilize.

One possible interpretation is that what is conventionally perceived as ``best practice'' code may not fully align with the LLMs' internalized representation of best practices. 
Another interpretation is that LLMs employ an alternative, model-specific notion of code quality that extends beyond established human conventions. 
Importantly, the diminishing number of code changes in subsequent iterations indicates that LLMs converge toward stability unless explicitly prompted to continue applying targeted changes (e.g., encouraged renamings). This raises the question of whether defining a convergence criterion is necessary at all, or whether such a criterion could effectively minimize or even eliminate the large-scale alterations observed in the initial iterations.

Another striking observation is that, in the absence of explicit prompting, LLMs consistently remove comments from the code. While this behavior might streamline the code’s appearance, it risks impairing readability for human developers and erases potentially valuable contextual information. Such comments may serve to document design rationales, highlight critical sections of the code, or record bug fixes and other implementation-specific considerations that would otherwise be lost. For example, comments that extend beyond mere functional descriptions (e.g., those addressing error resolution) exhibit a particularly strong correlation with successful bug fixing~\cite{song_empirical_2020}. On the other hand, poorly written comments are more harmful than having no comments at all, as they provide software developers and maintainers with distorted and misleading information~\cite{zhao_survey_2020}. Consequently, removing comments that do not align with the code context may in fact improve readability rather than harm it~\cite{Abdelsalam:Comments}. However, substantiating such an effect would require either a dedicated qualitative study or the application of advanced natural language processing techniques capable of assessing the contextual appropriateness of code comments.


\subsection{Interpretation for Convergence Across Code Variants (RQ\textsubscript{2})}

The analysis of RQ\textsubscript{2} reveals an interesting pattern: Despite starting from variants with degraded readability features, such as meaningless naming or missing comments, most versions ultimately converge toward highly similar code after several refactoring iterations. 
Variants initialized with meaningless identifiers and comments indeed underwent a larger number of renaming operations, particularly in the early stages of refactoring. This aligns with prior findings demonstrating that LLMs are capable of deriving clearer and more meaningful variable names even when the code is difficult to read or intentionally obfuscated~\cite{depalma_exploring_2024}.

This behavior points to a broader implication: 
LLMs tend to systematically reduce structural and stylistic differences between code variants, producing what could be described as a homogenization effect. 
The high frequency of renamings in poorly chosen identifier sets suggests that LLMs possesses a relatively fine-grained understanding of what constitutes ``good'' variable naming in the given context. While the present analysis cannot fully determine whether the chosen names optimally enhance readability or semantic clarity, our spot checks indicate that the replacements were context-appropriate and consistently expressed the intended function of the variables. 
A qualitative follow-up study would be necessary to more rigorously assess whether the new identifiers improve readability for human developers.

Interestingly, the overall convergence stands in contrast to initial expectations. One might have hypothesized that, given the vast space of possible code formulations, variants would diverge into increasingly distinct versions over multiple iterations. 
Instead, the opposite is true in that, across different starting conditions, the end versions gravitated toward a common form. 
This outcome may, in fact, be considered a best-case scenario, as it demonstrates that LLMs do not simply generate arbitrary code changes, but rather follows a discernible direction in refactoring code toward internally consistent and arguably more standardized solutions.


\subsection{Interpretation for Impact of Explicitly Emphasizing Refactoring Aspects (RQ\textsubscript{3})}

Our findings highlight that explicitly emphasizing particular readability aspect substantially shaped the refactoring trajectory. Naming-focused prompts consistently induced renaming, whereas comment-focused prompts produced a more stabilized outcome. This pattern resonates with prior work on refactoring interactions, which also observed that more specific prompts yield more targeted and effective model responses~\cite{alomar_how_2024, guo_exploring_2024}.

A closer comparison underscores two complementary insights. First, when guided by naming-focused prompts, LLMs engaged less in structural reorganization or code expansion. Auxiliary methods were seldom introduced, and control flow restructuring was minimized. This constrained focus suggests that explicit task framing steers LLMs away from their otherwise more creative or exploratory refactorings, a tendency that is consistent with recent findings showing that prompts and role-play settings significantly influence the creativity of LLMs~\cite{zhao_assessing_2025}. However, the same condition also produced a recurring pattern of continous renaming. The absence of a strict convergence criterion with our implemented prompts appears to allow continuous oscillation around variable names rather than stabilization once an arguably optimal naming scheme is reached.

By contrast, comment-focused prompts exhibited markedly different dynamics. Here, a trend to convergence occurred even in the absence of an explicit convergence criterion. LLMs appeared to ``recognize'' that additional comment insertions would not further improve readability, thereby stabilizing after only minor adjustments. While it remains unclear whether this reflects genuine semantic reasoning or merely a reduction in change tendencies under comment-emphasis, the effect is notable: Targeted prompting does not inherently prevent convergence, but its outcome depends on the nature of the targeted readability aspect.
Interestingly, despite the prompt's explicit focus on comments, a substantial number of renaming operations still occurred particularly in the variant with meaningless identifiers, indicating that LLMs might recognize the quality of identifier names and prioritizes it even under a prompt targeted at another readability aspect.

\section{Robustness Analyses, Open Issues, and Threats To Validity}
\label{ch:threats}

In this section, we present two follow-up experiments that scrutinize the robustness of our results. Further, we discuss open issues, which set the stage for identifying future research directions, and present major threats to validity.

\subsection{Robustness to Functional Correctness}

One limitation of our main experiment is that we did not evaluate the executability or functional correctness of the generated code. This omission implies a potential threat: One the one hand, LLMs may introduce errors when refactoring code. On the other hand, the probability of introducing such errors may increase with the number of refactoring iterations applied to a code snippet. When only concentrating on the structural changes between versions, we cannot rule out the possibility that some refactorings reduced functional correctness while still appearing to improve readability or structure.

To address this potential issue, we have analyzed all original and generated code snippets by executing the existing test cases. The 230 snippets we included in our main experiment are from a GitHub repository, which at the time of conducting the analysis, provided usable unit tests for 154 of these snippets.

\subsubsection{Methodology}

While executing the unit tests on the original version is trivial, for the \VarMeaningless~variant and many subsequent iterations of the LLM, this is challenging. We often cannot run the tests directly on these snippet versions, since either we renamed the methods during the variant creation or the LLM applied renamings and restructuring during the iterative refactorings. To overcome such structural renamings that are independent of the functional correctness, we have implemented a simple automated heuristic that reverses name changes to fit to the naming scheme of the given test case as best as possible. 

With this setup, we ran all usable unit tests. It is important to highlight that the LLM was not provided with the test cases during the refactoring, which led to a situation that sometimes test cases were failing that were only due to changes of superficial nature (e.g., changing the phrasing of the message of a thrown exception---breaking the existing unit test that was checking for a specific wording). Similarly, in a few cases, the LLM applied reasonable generalizations that made test cases fail (e.g., a function originally returned an \texttt{ArrayList} for which the test case was specifically testing for, but the LLM generalized it to return a \texttt{List}). Thus, we manually reviewed all failing test cases for whether they were truly breaking the underlying functionality or rather only changed something superficial.

\subsubsection{Results}

\begin{table}
    \centering
    \caption{Average passing rate of provided unit tests on different variant and versions.}
    \label{tab:robustness-passing-rate}
    \begin{tabular}{lrrrrrr}
    \toprule
    Variant     & v0 & v1 & v2 & v3 & v4 & v5 \\
    \midrule
    \VarOriginal    & 99.49\% & 98.48\% & 98.27\% & 98.32\% & 98.17\% & 98.24\%\\
    \VarMeaningless & 99.27\% & 98.57\% & 98.03\% & 97.79\% & 97.76\% & 97.70\%\\
    \VarNoComment  & 99.49\% & 98.53\% & 98.15\% & 98.17\% & 97.75\% & 97.80\%\\
    \bottomrule
    \end{tabular}
\end{table}

After the manual review and a correction in a handful of failing unit tests, we have collected the results of 59\,489 unit tests, of which 58\,419 passed and 1\,070 unit tests failed either due to a semantic change or a compile error. In~Table~\ref{tab:robustness-passing-rate}, we provide the average passing rate for the unit tests on different variant and versions. Notably, even the original version (v0) before passing it to the LLM did not yield a 100\% pass rate, which appears to be due to our different setup breaking some unit tests (e.g., hard-coded \texttt{\textbackslash n} line breaks in the unit tests). Nevertheless, the average passing rate drops only very slightly with each iteration, despite a prompt that specifically requested a refactoring targeting readability.

Our manual inspection of failed test cases revealed that typically a refactoring does not break the entire functionality, but creates an issue in some edge cases (e.g., empty input). Occasionally, a subsequent iteration fixed such issue. Nevertheless, this is an important lesson that seemingly innocent refactoring can actually introduce bugs that may not be obvious without detailed verification by the programmer.

\RQAnswerW{0.85}{\makecell{Functional\\Correctness}}{Based on the provided test suite, we executed 59\,489 unit tests on 154 snippets. While we accepted some superficial structural changes and generalizations, refactoring by the LLM led only to a slight drop ($<1\%$) in functional correctness with each iteration. Thus, our main results regarding the readability refactoring in RQ\textsubscript{1--3} are unaffected by the minuscule changes in functional correctness.}

\subsection{Robustness to Novel Code}

Another robustness concern arises from our snippet dataset. Since we relied on code snippets from a publicly available repository, which additionally covers fairly common algorithms, it is likely that parts of this material were included in GPT 5.1's training data. While this may have influenced the results of the specific refactorings proposed by the LLM, it does not diminish the value of our framework, which systematically captures and analyzes changes regardless of the source. In principle, the same pipeline can be applied to proprietary code that is unlikely to have been part of the training data, which was the goal of our second follow-up experiment.

\subsubsection{Methodology}

Selecting the right dataset, which has sufficient breadth, depth, and code quality, was a major challenge for our main experiment. When adding the constraint that it ideally was not part of the training data for GPT 5.1, and can be shared publicly for transparency, the options are extremely limited.

To find code snippets that fulfill all of the described criteria, and are similar to the code snippets of the main experiment, we decided on the following approach. We queried the GitHub API for Java projects, which were created in 2025 or later (i.e., after the training data cutoff date for GPT 5.1 in September 2024). We requested the top 1\,000 repositories that are available under an MIT license and sorted by the number of stars (as a rough indictor of code quality).

For each of the 1\,000 repositories, we crawled all containing Java files and automatically tested for the following criteria:

\begin{enumerate}
    \item The file is 50--200 LOC (same as main experiment).
    \item The file contains multi-line and inline Java comments.
    \item The comments are in English.
    \item We accept a maximum of 5 files from a single repository to ensure diversity in the snippet content.
\end{enumerate}

This yielded us 49 candidate snippets, which we manually reviewed to be of sufficient quality. We excluded files that only contained Java tests, code stubs without meaningful logic, when a copyright notice indicated a creation year before 2025, or when comments contained language other than English (which was missed by the automated process). From scraping 1\,000 repositories, this left us with 17 Java files that fulfilled all criteria to be code snippets for this follow-up experiment. As the final step, we conducted the exact same analysis approach as in the main experiment.

Note that, while this follow-up experiment cannot be exactly replicated since the GitHub API responses (and its underlying data) constantly change, we provide the full script for this analysis as well as all response data (including the excluded Java files) in our online replication package.

\subsubsection{Results}

\begin{figure}[ht!]
	\centering
	\includegraphics[width=0.8\linewidth, keepaspectratio]{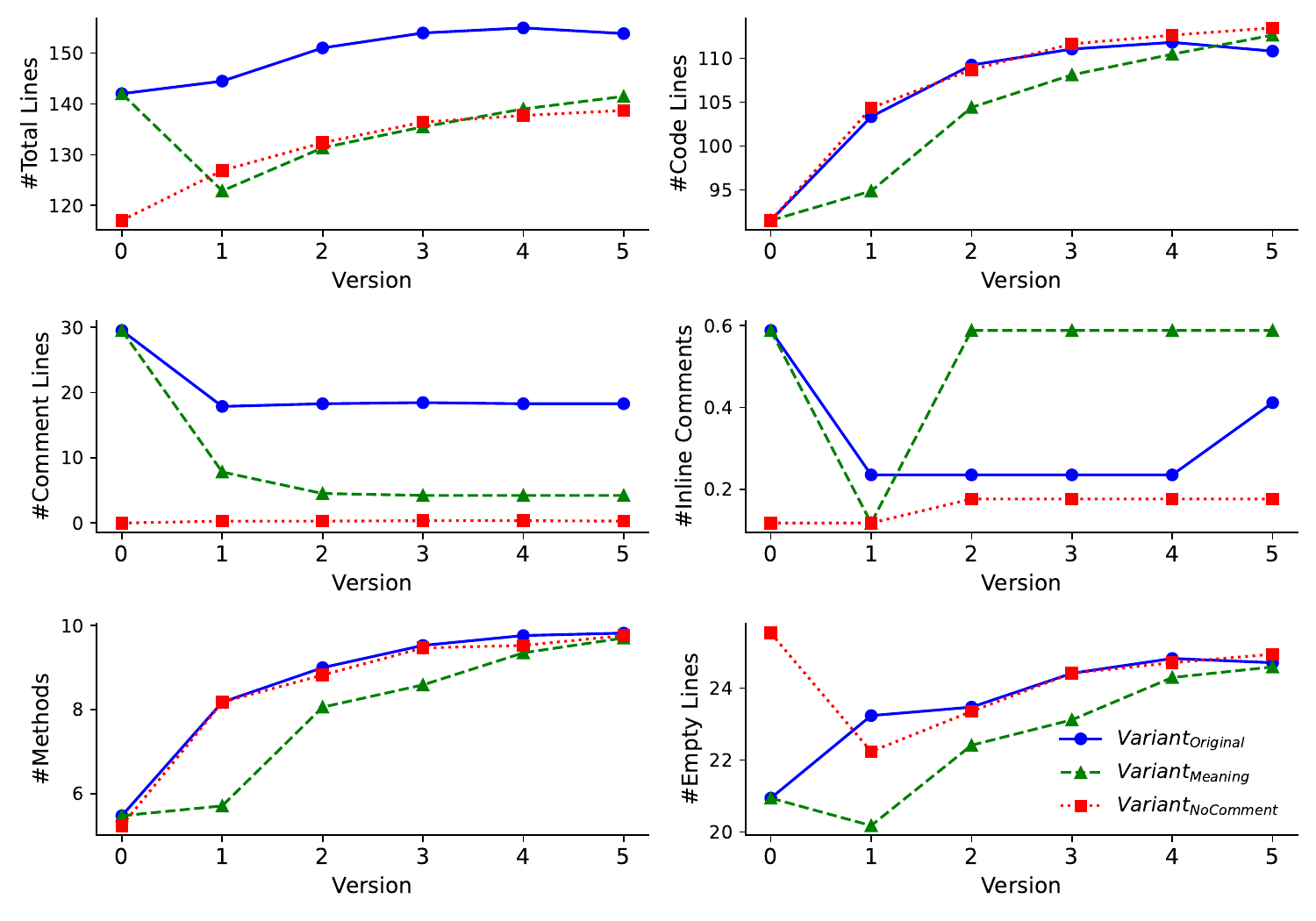}
	\caption{Average absolute metrics across refactoring versions for all three variants with Prompt\textsubscript{General} for our novel dataset.}
    \Description{TODO}
	\label{fig:robustness_absolute_values_across_all-snippets_KF0_only}
\end{figure}

\begin{figure}[ht!]
	\centering
	\includegraphics[width=0.85\linewidth, keepaspectratio]{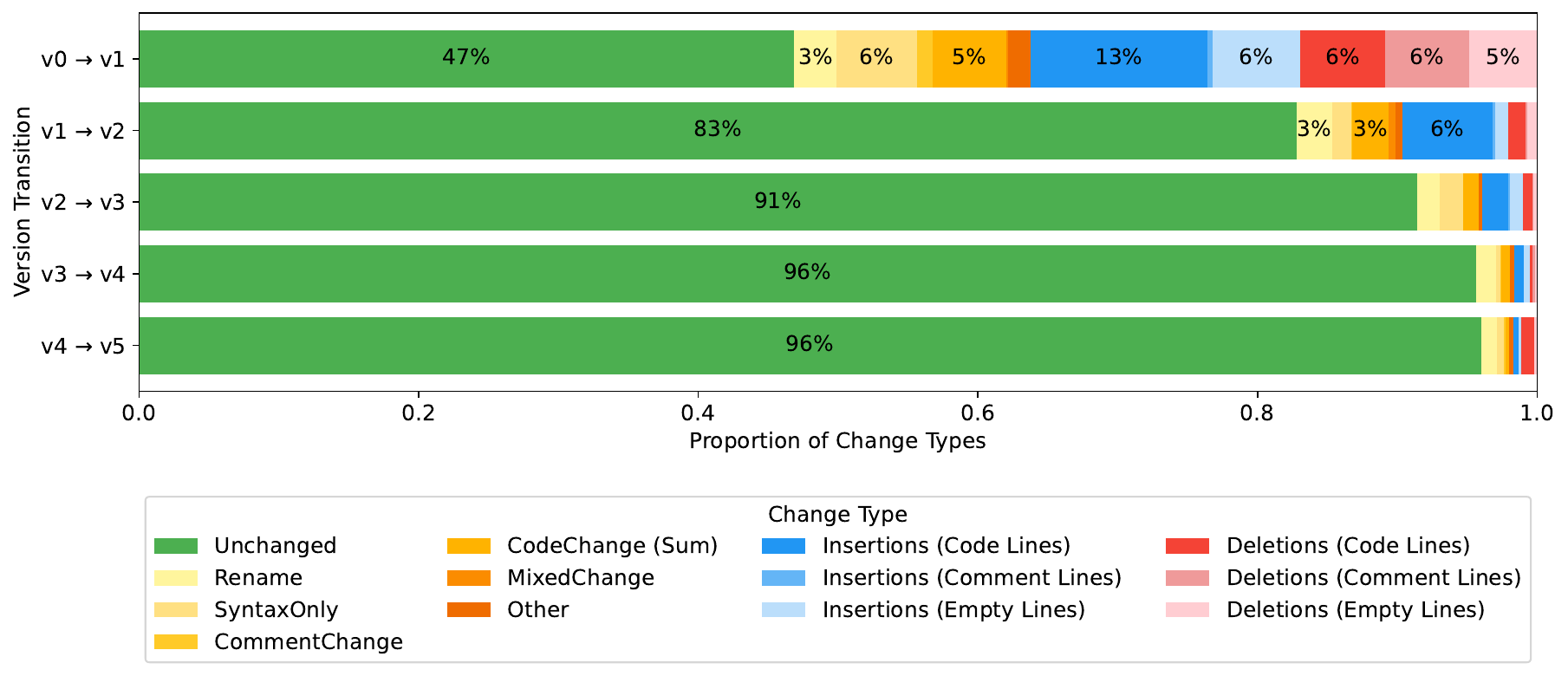}
	\caption{Distribution of code changes across iterations (\VarOriginal, Prompt\textsubscript{General}) for our novel dataset.}
    \Description{TODO}
	\label{fig:robustness_stacked_bar_plot_horizontal_main-exp_KF0_pKF0_detailed}
\end{figure}

To keep this manuscript concise, we focus on the most important results of our robustness analysis. Overall, the results for RQ\textsubscript{1} for novel code are comparable to our main experiment. Specifically, the absolute metrics show a similar trend across the three versions, as we visualize in Figure~\ref{fig:robustness_absolute_values_across_all-snippets_KF0_only}. In the same vein, we observe a large change in the first refactoring (v0 \textrightarrow v1) and subsequent smaller changes (cf.~Figure~\ref{fig:robustness_stacked_bar_plot_horizontal_main-exp_KF0_pKF0_detailed}).

\begin{figure}[ht!]
    \begin{minipage}{0.48\textwidth}
        \centering
        \includegraphics[width=\textwidth, keepaspectratio]{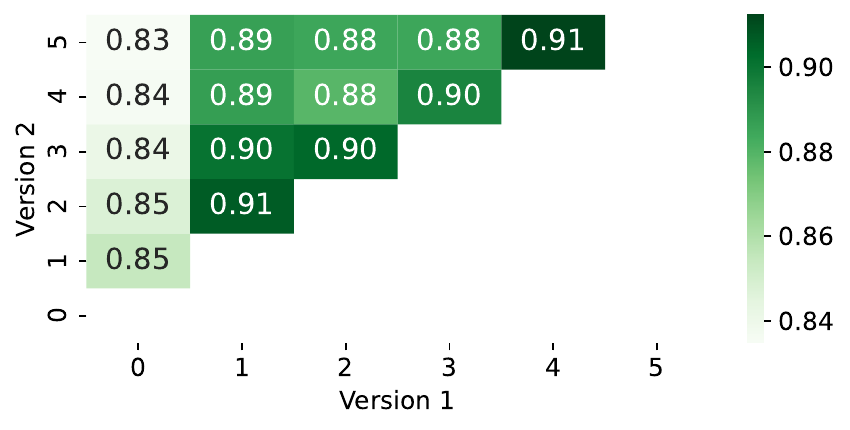}
        \caption{Similarity across iterations (\VarMeaningless, Prompt\textsubscript{Meaning}) for our novel dataset.}
        \Description{TODO}
        \label{fig:robustness_avg_simscore_heatmap_KF1_pKF1}
    \end{minipage}
    \hfill
    \begin{minipage}{0.48\textwidth}
        \centering
    	\includegraphics[width=\textwidth, keepaspectratio]{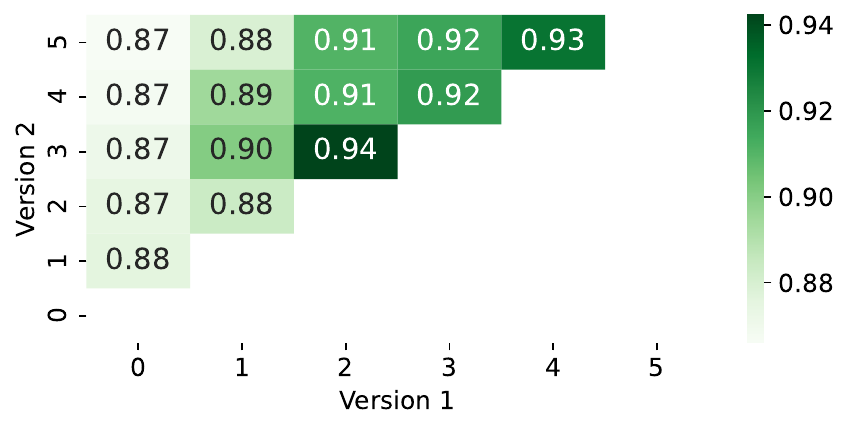}
    	\caption{Similarity across iterations (\VarNoComment, Prompt\textsubscript{Comments}) for our novel dataset.}
        \Description{TODO}
    	\label{fig:robustness_avg_simscore_heatmap_KF2_pKF2}
    \end{minipage}
\end{figure}

Regarding RQ\textsubscript{2} and RQ\textsubscript{3}, we also observe similar trends. As an example, we show the similarity scores between the \VarMeaningless~ (Figure~\ref{fig:robustness_avg_simscore_heatmap_KF1_pKF1}) and \VarNoComment~ (Figure~\ref{fig:robustness_avg_simscore_heatmap_KF2_pKF2}), which indicate a similar trend over the five refactoring iterations. We provide the full results with all generated figures in our replication package.

\RQAnswerW{0.92}{\makecell{Novel\\Code}}{Our analysis on code snippets that are novel for the LLM indicates that our findings from the main experiment are robust to a broader range of scenarios. While we still cannot rule out all potential biases, it appears that refactoring by the LLM is driven by a more fundamental understanding of code, and not just memorization of the training data.}

\subsection{Open Issues}

\subsubsection{Convergence vs. Oscillation}

Our results demonstrate that iterative refactorings exhibit a clear pattern of early restructuring followed by stabilization. This supports the findings by Liu et al.~\cite{liu_iterative_2025}, who found that the majority of code improvements occur in early iterations.  
In line with the open question of whether LLMs converge toward higher-quality solutions, we find that the LLM consistently normalizes structurally divergent inputs, even when identifiers are meaningless or comments removed. However, convergence remains partial: Similarity scores often plateau below 1.0, and micro-changes persist across iterations, indicating that the LLM occasionally engages in over-refactoring and that it might benefit from a principled stopping criterion.

An additional dimension that may influence convergence is the availability of context. Given that API requests are stateless, the back-and-forth changes cannot be explained by the LLM retaining information about previous snippet versions. Rather, the observed oscillation appears to arise from internal distribution of the LLM over plausible identifier names, which leads it to revert to familiar candidates in the absence of persistent contextual anchoring.

Future research shall therefore examine whether maintaining context across refactoring iterations would alter convergence dynamics, potentially reducing oscillatory behavior. Nevertheless, our findings already indicate that in a stateless setup, the LLM sometimes oscillates within a narrow set of highly similar identifier names when repeatedly prompted to improve naming. This suggests that, without contextual anchoring, LLMs may converge only locally and remain vulnerable to back-and-forth code changes in semantically equivalent but stylistically similar solutions.

\subsubsection{Decomposition of Refactoring Types}

By analyzing insertions, deletions, and different change types of refactorings over time, our main experiment provides the layered characterization of iterative refactoring that was previously missing. Our results reveal that early iterations are dominated by structural refactoring (e.g., comment deletions, renaming, or code changes), while later iterations narrow down to minor surface-level edits. This raises further questions regarding the nature of these structural refactoring. Which code locations are particularly frequent or rare targets of change? 
Which node types of an AST are most commonly affected? 
Which change types are especially associated with specific AST node categories? Moreover, which change types tend to co-occur, and which are largely mutually exclusive? 
Another important aspect is whether code size plays a role. For instance, are certain change types applied more frequently in larger snippets compared to smaller ones?

\begin{figure}[ht!]
	\centering
	\includegraphics[width=0.8\linewidth, keepaspectratio]{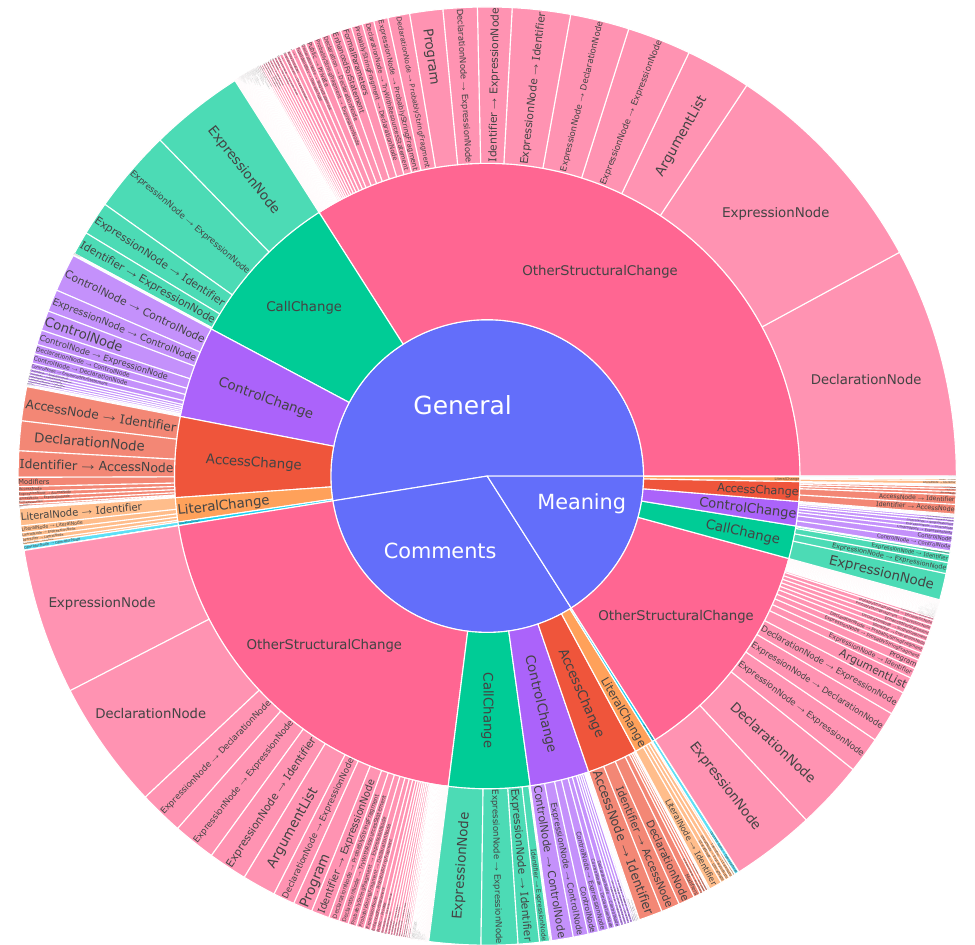}
	\caption{Sunburst diagram showing the distribution of AST node types affected by code changes, grouped by prompt (center), change type (middle layer), and node category (outer layer)}
    \Description{TODO}
	\label{fig:sunburst-semantic-ast}
\end{figure}

These questions go beyond the scope of this work though. Nevertheless, the framework has been designed in such a way that the necessary data (i.e., information on which AST node types are affected by a given change at the line level) are already partially collected during the analysis. We provide an exemplary overview of affected AST node types per code change type in Figure~\ref{fig:sunburst-semantic-ast}. The figure presents a hierarchical distribution of changes across prompts, code change types, and affected AST node categories. It shows that a substantial portion of current detections are still grouped under \textit{OtherStructuralChange}, reflecting the early stage of implementation for code change classification. Nonetheless, clear patterns already emerge for implemented categories such as \textit{CallChange} or \textit{ControlChange}, which frequently affect specific node types such as \texttt{ExpressionNode} or \texttt{ControlNode}. This structure lays the groundwork for more fine-grained analyses, including the co-occurrence of change types and the identification of structural hotspots within code. Understanding which AST node types are affected by specific change types can shed light on common refactoring strategies, developer behavior, and code evolution over time. This knowledge could inform the development of smarter tooling for code review, automated refactoring, and the evaluation of code-generating models. Moreover, it provides a foundation for more nuanced comparisons between different kinds of prompts or user strategies.

\subsection{Further Threats to Validity}

Although the experiment was carefully designed, certain limitations remain beyond the already covered aspects of the robustness analyses that may affect the interpretation and generalizability of the results. Following established guidelines in empirical software engineering~\cite{Jedlitschka2008}, we discuss threats to validity along four dimensions: Internal validity (factors that may have influenced the correctness of our measurements), external validity (the extent to which our findings can be generalized), conclusion validity (the robustness of the inferences drawn from the data), and construct validity (the adequacy of our operationalization of the studied concepts).

\subsubsection{Internal Validity}

The accuracy of our results depends directly on the performance of the matching process implemented in our parser. Correctly classifying actual code changes is a non-trivial task, and while the parser was iteratively improved, the dataset used during development naturally could not cover all possible cases of code line pairs that may occur in a diff. It is therefore possible that some lines were mismatched, left unmatched, or otherwise incorrectly processed. Such errors may have introduced noise into the analysis, potentially affecting the specific precision of our measurements. However, since our analysis rather focus on the big picture, such as the convergence trends, across many snippets potential line mismatches in a few snippets do not affect the overall results.

Beyond the matching process itself, further limitations arise from the design of our change-type classification. One notable example concerns the handling of rename operations. Rename operations were not normalized in our analysis. This means that if a variable occurred multiple times in the code, each instance was counted as a separate renaming operation. While the parser could in principle be extended to consolidate such cases into a single renaming event, this is a non-trivial enhancement. At the same time, reporting absolute counts is informative, as renaming a variable with many occurrences can have greater practical impact than changing a variable used only once. Thus, although this design choice should be kept in mind, it also does not fundamentally distort the overall results. A more critical question, however, would be whether the model consistently renames all instances of a given variable. Inconsistent renaming would not only produce misleading counts but could also render the code dysfunctional or subtly alter its behavior. While such issues may show up in our robustness analysis regarding functional correctness, a direct exploration of this aspect remains open. This limitation should be kept in mind when interpreting the relative weight of renaming operations in the overall results.

\subsubsection{External Validity}

The generalizability of our findings is limited by the fact that we relied on a single model throughout the main experiment. Moreover, we did not employ the most recent version at the time of writing~(\texttt{gpt5.2}), as it became available only after the data collection had been completed.
Our work did not focus on carving out differences between different LLMs, so we cannot claim with certainty that our results all fully extend to other LLMs. Since architecturally LLMs are fairly similar, we would expect these results to be comparable. Our anecdotal evidence points toward a stable effect: An early version of this work was a master's thesis using \texttt{gpt4.1-mini}, which showed the same trends~\cite{HessThesis}. When preparing the presented experiment, we initially tested the technical setup (for financial and time reasons) with \texttt{gpt5.1-mini}. Our impressions across these three results are that the larger models are more ``opinionated'', such that the absolute change numbers are a bit larger (e.g., more renamings across all five iterations), but across all three models, the main results were the same. This also highlights that the methodological pipeline itself is not bound to any specific model. If an API is available, only minimal adjustments to the request format and conversion of the response are required, which allows the approach to be readily transferred to different LLMs.

\subsubsection{Conclusion Validity}

By averaging results across a large number of snippets and variants, important details may be obscured. For example, back-and-forth changes can occur in individual refactoring sequences, yet their impact may be diluted in the aggregate analysis. As a consequence, such phenomena might not strongly affect overall similarity scores, even though they represent meaningful dynamics at the micro-level. Capturing these nuances would require complementary qualitative analyses, which were beyond the scope of this experiment.

Another threat arises from the design of our similarity measure. Our similarity score does not account for the change types of insertions and deletions, and therefore does not theoretically capture the overall similarity between entire snippets. Instead, it reflects only the similarity of the changed code segments between two successive versions. This design choice could, in principle, distort the measurement of convergence. However, when considering the distribution of change types across iterations, it becomes evident that the proportion of insertions and deletions decreases sharply with higher version numbers and is negligible in later stages. As a result, the similarity score remains a representative proxy for overall similarity in the context of this experiment.

Insertions and deletions were also not explicitly represented as distinct change types in our classification scheme. Similar to the design of the average similarity score, this choice could in principle limit the completeness of the analysis. However, the distribution of change types across iterations shows that insertions and deletions occur only rarely in later versions and therefore do not substantially affect the main findings. For the sake of completeness, though, future work could extend the classification to also capture code growth or reduction explicitly, for example, by reflecting the expansion or contraction of comment sections rather than focusing solely on changes to existing comments.

A further concern relates to the inherent non-determinism of LLM outputs. Ideally, the experiment should be repeated multiple times to examine whether the overall outcomes remain stable across different runs. Such replication would provide stronger evidence regarding the reliability of the observed patterns. In our experiment, however, we followed current best practices by setting the generation parameter of temperature to $0$, which is commonly used to minimize randomness. While this increases reproducibility to some extent, it does not fully eliminate the possibility that repeated runs could yield slightly different results.

\subsubsection{Construct Validity}

A limitation of our experiment lies in the fact that we did not directly measure whether the resulting code was indeed more readable. While we frequently observed indications of improved readability during development, these impressions remain anecdotal rather than systematically validated. Nevertheless, what makes code ``readable'' is still ongoing research~\cite{Piantadosi2020}. 

An additional approach could have involved the use of static analysis tools to assess commonly established code quality issues, as demonstrated in related work~\cite{liu_iterative_2025}. Such tools might have been employed to first evaluate the ``best practice'' snippets and identify the improvements they suggest, thereby enabling a comparison with refactorings by the LLM. However, this approach has its own set of limitations regarding the accuracy of such tools---and whether they actually improve human readability at all. In our experiment, we qualitatively observed clear improvements: the LLM consistently produced more uniform and well-formatted code outputs and, in cases of variable renaming, was able to correctly infer and assign meaningful names based on context. Prior research has likewise demonstrated strong capability of LLMs to summarize code and capture contextual information~\cite{tian_is_2023}. In our experiment, given the size of the dataset, we relied on sample-based qualitative checks rather than systematic validation. This decision also reflected our primary focus on building a framework to measure code changes across arbitrary iterations of LLM-optimized code. Ultimately, determining whether one snippet is more readable than another remains inherently subjective however, underscoring the need for a larger-scale qualitative study to rigorously assess improvements or regressions in code readability.

\section{Conclusion}
\label{ch:conclusion}

This work investigated the capability and consistency of LLMs to iteratively refactor source code regarding its readability. We conducted a large-scale main experiment using OpenAI’s GPT5.1 with 230 Java snippets, each systematically varied and
refactored with respect to readability across five iterations under three different prompting strategies. We analyzed with a particular focus on convergence and the role of prompting strategies. The results for source code that already follows best practices show that LLMs typically preserved structure but introduced minor, sometimes unnecessary, code changes before stabilizing. Across variants which violate typical conventions for high quality code, such as meaningful identifiers or comments, iterative refactoring by the LLM still converged toward highly similar final versions, suggesting a normalization effect in which LLMs align diverse inputs with implicit coding conventions. Prompt design proved critical: Emphasizing variable naming frequently led to oscillatory changes, whereas prompts stressing comments supported faster stabilization. A follow-up experiment showed that these results are unaffected by the potential of breaking semantic functionality during the refactoring. Another follow-up experiment demonstrated that the results hold even for novel code.

Taken together, our work provides an empirical foundation for assessing the reliability of LLM-assisted code improvement.
Future research should extend this work along several dimensions: Exploring larger and more diverse code bases or systematically varying prompt strategies and contextual anchoring. By addressing these aspects, subsequent studies can further clarify the conditions under which LLM-driven refactorings contribute meaningfully to code quality. In doing so, they will bring us closer to harnessing the full potential of large language models as partners in software development.

\begin{acks}
The work is supported by ERC Advanced Grant ``Brains On Code – A Neuroscientific Foundation of Program Comprehension'' 101052182.
\end{acks}

\bibliographystyle{ACM-Reference-Format} 
\bibliography{References}

\end{document}